\documentclass[aps,prb,preprint,groupedaddress,showpacs,floatfix]{revtex4}
\usepackage{graphicx}
\usepackage{longtable}
\usepackage{bm}

\begin{document}

\preprint{}

\title{Magnetic structure of free cobalt clusters studied with Stern-Gerlach deflection experiments}

\author{F.W. Payne, Wei Jiang, J.W. Emmert, Jun Deng, and L.A. Bloomfield}
\email{lab3e@Virginia.edu}
\affiliation{Department of Physics, University of Virginia, 
Charlottesville, Va. 22904}

\date{\today}

\begin{abstract}
We have studied the magnetic properties of free cobalt clusters in two semi-independent Stern-Gerlach deflection experiments at temperatures between 60 and 307~K. We find that clusters consisting of 13 to 200 cobalt atoms exhibit behavior that is entirely consistent with superparamagnetism, though complicated by finite-system fluctuations in cluster temperature. By fitting the data to the Langevin function, we report magnetic moments per atom for each cobalt cluster size and compare the results of our two measurements and all those performed previously. In addition to a gradual decrease in moment per atom with increasing size, there are oscillations that appear to be caused by geometrical shell structure. We discuss our observations in light of the two competing models for Langevin-like magnetization behavior in free clusters, superparamagnetism and adiabatic magnetization, and conclude that the evidence strongly supports the superparamagnetic model.\end{abstract}


\maketitle

\section{Introduction} 
\label{Introduction}
Interest in small magnetic systems predates the emerging field of nanoscience by several decades.~\cite{neel1949,bean1955} Magnetism in small systems is effectively the frontier between atomic magnetism, surface magnetism, and bulk magnetism. Because of its strategic position, small system magnetism has both theoretical importance in elucidating how structural and electronic order evolve from one dimensional extreme to the other and practical importance in such technologies as magnetic data storage and high-performance magnetic materials.

Magnetism in small and low-dimensional systems has been studied since the 1950s using powders, granular metals, bulk surfaces, and supported films and particles.~\cite{bean1955,jacobs1963} Phenomena observed in these contexts included superparamagnetism and surface-enhanced magnetism. Superparamagnetism occurs when the overall moment of a small magnetically ordered particle becomes thermally decoupled from its lattice so that it can respond paramagnetically to an applied magnetic field.\cite{neel1949,bean1955,bean1959,brown1959} Surface-enhanced magnetism results when the decreased coordination number and increased density of states at a surface reduces the quenching of spin and orbital magnetism relative to what occurs in the bulk.~\cite{liu1991,dunlap1991}

More recently, however, the magnetic properties of atomic clusters have been studied in the isolation of vacuum using molecular beam techniques.~\cite{cox1985,deheer1990,bucher1991,douglass1992,douglass1993,cox1993,bucher1993,billas1993,cox1994,billas1994,apsel1996,knickelbein2001,xu2005,knickelbein2006} That isolation has at least three interesting consequences: it frees the clusters from all chemical interactions at their surfaces, it decouples them from external heat baths, and it emphasizes their conserved quantities. As we will discuss, those isolation effects both simplify and complicate the observed behaviors of magnetic clusters in a beam.

One might expect that a beam of magnetic molecules of spin $S$ would, upon passing though a gradient magnetic field, split into $2S+1$ beamlets. This behavior is indeed the high-field case for atoms and small molecules.~\cite{Amirav1981,Kuebler1988} One might also argue that the $N$ atomic moments $\mu$ in a ferromagnetic particle would couple together and orient themselves along the applied magnetic field, and that the particle as a whole would then accelerate and deflect toward high field in response to the force ${\bm \nabla}(N{\bm {\mu\cdot B}})$. That behavior is the high-field, low-temperature limit for single-domain condensed matter particles. The clusters we study are intermediate in size and we study them at moderate temperatures and magnetic fields. Anticipating their behaviors is therefore not so simple.

The first magnetic deflection experiments on free cobalt clusters observed that these clusters always deflect toward strong field by an amount that increases as the applied magnetic field increases, as the relevant temperature decreases, and as the number of atoms in the cluster increases.~\cite{douglass1993} In the present measurements, we again observe deflections that increase with increasing magnetic field and decreasing temperature, and that are entirely consistent with superparamagnetic behavior in which each cluster acts as its own heat bath at its vibrational temperature.\cite{khanna1991} The influence of cluster size, however, is more complicated than we observed originally. We now find that the each cobalt cluster's magnetic moment depends subtly on the number and arrangement of its atoms.

In superparamagnetism, the effective magnetic moment per atom $\mu_{\rm eff}$ is reduced from the true moment per atom $\mu$ by the Langevin function $\cal L$
\begin{eqnarray}
\mu_{\rm eff} & = & \mu{\cal L}(N\mu B/k_BT_{\rm vib})\label{eq:langevin}\\ & = & \mu[\coth(N\mu B/k_BT_{\rm vib})-(k_BT_{\rm vib}/N\mu B)]\nonumber
\end{eqnarray}
\noindent where $N$ is the number of atoms in the particle, $B$ is the applied magnetic field, $k_B$ is Boltzmann's constant, and $T_{\rm vib}$ is the vibrational temperature. This reduction occurs when thermal agitation decouples the particle's magnetic moment from its lattice and causes that moment to explore all possible orientations in rapid succession. Since the orientation fluctuations occur on a subnanosecond timescale,\cite{neel1949,brown1959,khanna1991} a millisecond-timescale measurement of the particle's magnetic momentum per atom will yield the time-averaged projection of ${\bm \mu}$ onto the applied magnetic field axis ${\bm {\hat B}}$---that is, the experiment will measure $\mu_{\rm eff}$. This experimental moment per atom increases linearly with applied magnetic field for weak fields, but eventually saturates at $\mu$.

Equation~(\ref{eq:langevin}) assumes that the total moment N$\mu$ is large enough to be treated classically, otherwise the Brillion function must replace the Langevin function. It also assumes that all orientations of the moment relative to the particle lattice are energetically equivalent, that the measurement time is long compared to the thermal fluctuation time so that the particle thoroughly explores the Boltzmann distribution during the measurement, and that the particle is coupled to an infinite heat bath at temperature $T_{\rm vib}$.~\cite{knickelbein2004}

While free clusters cannot meet these assumptions perfectly, there is considerable evidence that Eq.~(\ref{eq:langevin}) is valid for cobalt and most other magnetic clusters over a considerable range of experimental conditions. We note here that an adiabatic mechanism of magnetization has recently been proposed.~\cite{xu2005} That alternative theory arose as an explanation for the broadening of the cluster beam as it deflects. Interestingly, the adiabatic mechanism reportedly gives the same Langevin-like reduction in measured magnetic moment that is predicted by superparamagnetic theory. We will discuss these two competing explanations later in this paper. For the present, we will simply interpret our experimental data using Eq.~(\ref{eq:langevin}) in order to obtained values for $\mu$, the magnetic moments per atom of the cobalt clusters, and make occasional reference to superparamagnetism, the model that we ultimately conclude is responsible for our experimental observations.

What distinguishes the present study from the one performed more than a decade ago in our laboratory~\cite{douglass1993} is that we are now able to report $\mu$ for each cluster size individually. In the earlier study, our limited sample required us to average across cluster sizes. We therefore had to assume that $\mu$ was independent of cluster size, at least over the size range considered (40--200 atoms). In the present work, however, we have studied each cluster size individually, so that magnetic effects due to cluster size and structure can be recognized.

The present study consists of two semi-independent experiments, which we will refer to as E1 and E2. These experiments were conducted before (E1) and after (E2) a major renovation of the experimental apparatus. Most significantly, they were performed with radically different cluster sources and data acquisition protocols. Although E1 and E2 share the same gradient-field magnet and mass-spectrometer, their semi-independence makes them a useful test of our ability to control cluster temperature and therefore make meaningful measurements. That the results of these two studies are so similar provides considerable support for our assertion that we understand the temperatures of clusters in our beams.

\section{Experimental Procedure}
\label{Procedure}

In concept, the experiment is essentially the Stern-Gerlach deflection technique applied to cobalt clusters. We produce a narrowly collimated beam of temperature-controlled cobalt clusters, pass that beam through a gradient magnetic field, and measure the deflections and masses of the clusters at a distance downstream from the magnet. While we have described this technique previously~\cite{bucher1993}, we discuss it again here primarily to point out the differences between the two experiments, E1 and E2.

In both experiments, cobalt clusters were grown from atoms vaporized off a cobalt sample by the focused second-harmonic light (532 nm) of a pulsed Nd:YAG laser. The laser-produced plume of cobalt vapor was captured in a pulse of dense helium gas, where cluster growth and thermal equilibration occurred. This mixture of gas and clusters then became a cluster beam through a seeded supersonic expansion into vacuum.

In experiment E1, cobalt vapor and helium mixed in a cylindrical chamber approximately 8~mm in diameter and 17.5~mm long. Clusters grew in this ``waiting room'' and came gradually into thermal equilibrium with its walls before expanding into vacuum through the chamber's conical exit nozzle. In practice, however, clusters sprayed out of the nozzle continuously once cobalt atoms were injected into the helium pulse. We found that only those clusters that remained in the source as long as possible and didn't emerge from the chamber until it was almost empty had time to reach thermal equilibrium with the chamber walls. The measurements we report for E1 were all conducted with clusters that had resided in the waiting room long enough ($\ge$1.5~ms) to reach equilibrium. The range of cluster temperatures used in E1 was 63 K to 307 K, with 63 K being approximately the coldest source temperature that the closed-cycle helium refrigerator could maintain. 

In experiment E2, cobalt vapor and helium mixed and cooled in a cylindrical channel 2.5~mm in diameter and 150~mm long. This channel had approximately the same volume as the ``waiting room'' used in E1, but a much larger surface area. Cluster growth occurred near the start of this channel and the resulting clusters were thermally equilibrated in the channel's final 100~mm, which is temperature-controlled. The channel ends with a 1~mm cylindrical nozzle, through which the mixture of vapors underwent a seeded supersonic expansion to form a cluster beam. The increased surface to volume ratio of this source, coupled with the fact that even the fastest clusters take more time to get from the point of vaporization to the nozzle, means that a greater fraction of the clusters are in thermal equilibrium with the source upon exiting. All of the measurements we report for E2 were made using equilibrated cobalt clusters. The range of cluster temperatures used in E2 was 60 K to 100 K, again limited at the low end by the closed-cycle helium refrigerator. 

Both experiments shared the same beam shaping apparatus. The cluster beam travels through a skimmer and two narrow slits (0.4 mm wide by 2.5 mm high) that collimate the beam, and through a chopper wheel that slices it longitudinally. The chopper wheel performs two important functions: it allows us to select only those clusters that have come into thermal equilibrium with the source and it permits us to measure the velocity of the clusters. The wheel rotates at 180 Hz and is open to the cluster beam for $\sim$20 $\mu$s. After passing through these collimating slits and chopper, the cluster beam is a tightly grouped packet, narrow in all three spatial dimensions.

Clusters in this packet then pass through the gradient field magnet, where they accelerate in response to the force ${\bm \nabla}(N{\bm {\mu\cdot B}})$. Since the magnet's field gradient is perpendicular to the cluster beam axis, magnetic clusters are deflected, and it is that deflection that we subsequently measure.

The gradient field magnet is 250~mm long and its gradient is nearly constant in the region traversed by the cluster beam.~\cite{mccolm1964} We can vary the field gradient experienced by the clusters from 0~T/m to 360~T/m, while simultaneously varying the field at the position of the cluster packet from 0~T to 0.951~T.

After leaving the magnet, clusters pass through a $\sim$1~m field-free drift region and are then ionized by a spatially filtered pulse of laser light (193~nm, ArF Excimer) in a time-of-flight mass spectrometer. This narrow light beam propagates antiparallel to the cluster beam and ionizes only those clusters that have deflected a specific distance from the undeflected beam center. By recording the resulting mass spectra at a range of positions relative to the beam center, we obtain a complete deflection profile for each of the cluster sizes present in the beam.

In experiment E1, these deflection profiles were obtained one magnetic field at a time. With the source producing pulses of clusters at 20~Hz and the magnetic field and field gradient set to one value, we scanned the laser beam back and forth across the cluster beam and accumulated cluster deflection profiles for several hours. We then chose another field and field gradient and repeated this exercise. We periodically turned off the magnet to obtain a zero-field profile to establish the undeflected beam center.

Because the experimental apparatus is $\sim$3~m long and the temperature in the laboratory varies over the course of a day, there are small but unavoidable movements in the equipment. Furthermore, the intensity and direction of the cluster beam itself drifts with time as the sample ages and soot accumulates on the source's internal surfaces. These changes produce gradual shifts in the measured deflection profiles, and the one-field-at-a-time measurement approach used in experiment E1 makes it difficult to remove these shifts from the data.

Conversely, in experiment E2, through computer automation, we collected data randomly in both position and magnetic field, with the magnetic field switching between various values (including zero) and the ionizing laser shifting to a new randomly selected position a few times a minute. Although the source used in E2 produced pulses of clusters at 30~Hz, a single experimental run still spanned an entire day. Although the long duration of each experimental run in E2 may have increased the statistical uncertainty, experimental drift no longer contributed significantly to the systematic uncertainty.

\section{Results}
\label{Results}
\subsection{Residence Time Studies}

Because clusters spend $>$100~$\mu$s passing through the gradient field magnet, our experimental measurement timescale is long compared to the subnanosecond timescale of superparamagnetic fluctuations. Assuming that the cobalt clusters behave superparamagnetically in these experiments, the magnetic moment responsible for a cluster's deflection by the gradient field magnet is the time-averaged projection of that cluster's true moment onto the magnetic field. The experimentally measured moment per atom $\mu_{\rm expt}$ is thus the effective moment per atom $\mu_{\rm eff}$ given by Eq.~(\ref{eq:langevin}).

One of the best indications that cobalt clusters obey Eq.~(\ref{eq:langevin}) (and that they behave superparamagnetically in our experiments) is that $\mu_{\rm expt}$ has the correct functional dependence on applied magnetic field $B$ and cluster vibrational temperature $T_{\rm vib}$. In order to be sure that we know $T_{\rm vib}$, however, we must prove that the clusters have resided in the source long enough to reach thermal equilibrium with its walls. We can then use the source's wall temperature as $T_{\rm vib}$ in all the subsequent analysis.

Because knowing $T_{\rm vib}$ is so important, we regularly perform studies in which we measure $\mu_{\rm expt}$ as a function of residence time in the source. The residence time for a particular group of clusters is defined as the interval between the Nd:YAG laser pulse and the moment those clusters emerged from the nozzle into vacuum. We can determine the moment at which a particular group of clusters left the nozzle from the timings of the chopper wheel and the ionizing laser pulse and from the distances separating these components from the nozzle. Knowing when the clusters left the nozzle allows us to calculate their residence time to $\pm10~\mu$s.

\begin{figure}
\includegraphics{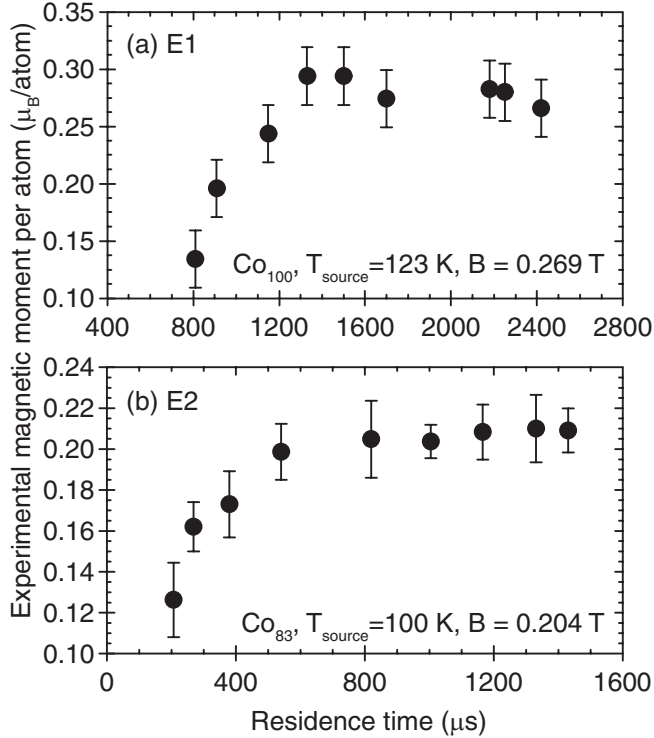}
\caption{\label{fig:restime} The experimental magnetic moment per atom $\mu_{\rm expt}$ initially increases with residence time in the source but eventually saturates when the cluster vibrational temperature $T_{\rm vib}$ reaches equilibrium with the source temperature. (a) is a saturation curve from experiment E1, taken for Co$_{100}$ at a source temperature of 123 K and a magnetic field of 0.269 T, and (b) is one from experiment E2, taken for Co$_{83}$ at a source temperature of 100 K and a magnetic field of 0.204 T.}
\end{figure}

As shown in Fig.~\ref{fig:restime}, we always observe an initial rise in $\mu_{\rm expt}$ with increasing residence time, followed by a long period of saturation in which additional residence time has no effect on the measured moment per atom. The timescales for the waiting room source used in E1 were longer than those for the channel source used in E2 because the waiting room had less surface area to cool the helium/cluster mixture and it emptied slowly through its small nozzle. The channel source blew its helium/cluster mixture quickly through its channel while providing more rapid cooling. Our final moment measurements were all conducted on clusters with residence times sufficient to ensure that they had reached thermal equilibrium with their sources and had vibrational temperatures $T_{\rm vib}$ equal to their source temperatures.

\subsection{Measured Magnetic Moments}

In both experiments, E1 and E2, we obtained deflection profiles for each cluster size at many temperatures $T_{\rm vib}$ and many magnetic fields $B$. We calculated a $\mu_{\rm expt}$ from each profile and then fit all the measurements for a given cluster size to Eq.~(\ref{eq:langevin}). The high quality of the fits demonstrates the validity of the Langevin magnetization relationship. The only free parameter in each fit is $\mu$, the true magnetic moment per atom for that cluster size.

\begin{figure*}
\includegraphics{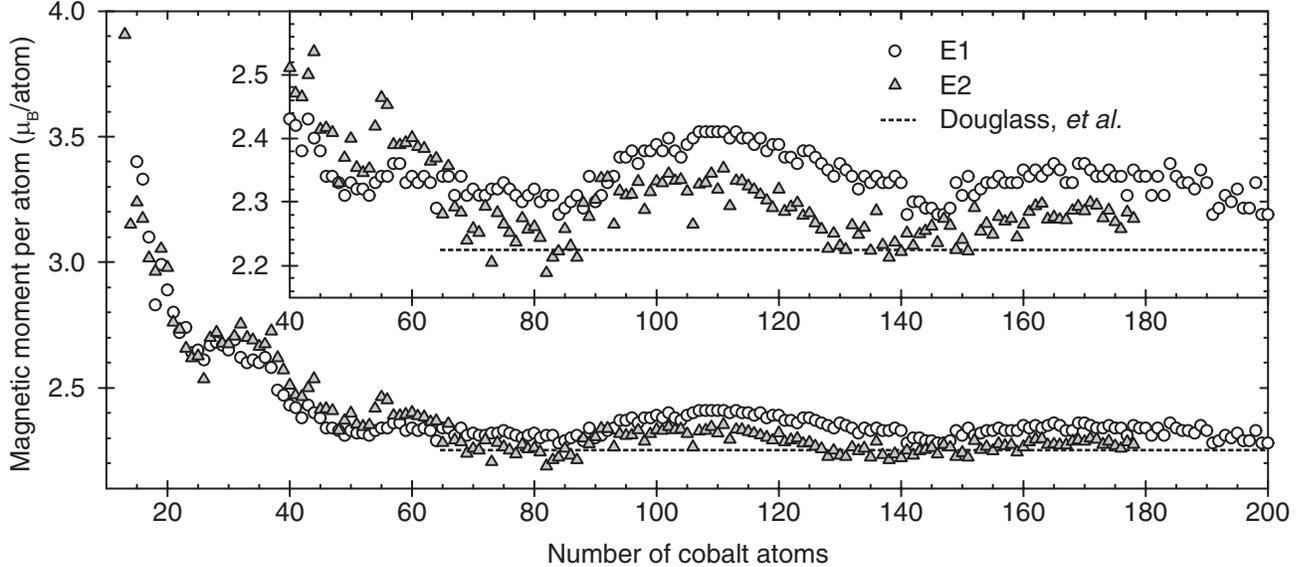}
\caption{\label{fig:moments} Magnetic moments per atom $\mu$ for cobalt clusters consisting of 13--200 atoms. Values obtained in Experiment E1 are represented by open circles and those obtained in Experiment E2 are represented by shaded triangles. The dashed line indicates the single average value measured for Co$_{65-215}$ by Douglass, {\it et al.} (Ref.~\onlinecite{douglass1993}).}
\end{figure*}

Figure~\ref{fig:moments} shows the values of $\mu$ obtained for each cluster size in the two experiments, E1 and E2, and those values are also listed in Table~\ref{tab:moments}. The listed values include only the statistical uncertainties, which are small and demonstrate the excellent agreement between Eq.~(\ref{eq:langevin}) and the experimental observations. The systematic uncertainties are larger and reflect limitations in the magnet calibration, position control of the cluster beam in the magnet, and the control of vibrational temperature. We estimate this systematic uncertainty at $\pm7\%$ in E1 and $\pm5\%$ in E2.

\section{Discussion}
\label{Discussion}

In addition to the two experiments reported here, there have been two other experimental studies of magnetism in free cobalt clusters published recently: a study by Xu {\it et al.}~\cite{xu2005} (hereafter referred to as Xu) covering the entire range of cluster sizes discussed in the present work and a study by Knickelbein\cite{knickelbein2006} focusing on clusters consisting of between 7 and 32 atoms. There are many important similarities and differences between the results of the four experiments.

One of their most notable and physically important similarities is that they all find enhanced magnetism, relative to that of bulk cobalt (1.7 $\mu_B$/atom), in all of the cobalt clusters that deflected toward strong field. This enhancement has been observed before~\cite{bucher1991,billas1993,douglass1993,billas1994,apsel1996} and anticipated theoretically due to the reduced dimensionality of the clusters.\cite{liu1991,dunlap1991} Their small size leads to an increase in the density of states at the Fermi level and to increased contributions from unquenched orbital and spin magnetism.~\cite{li1993,guirado2003}

\begin{figure}
\includegraphics{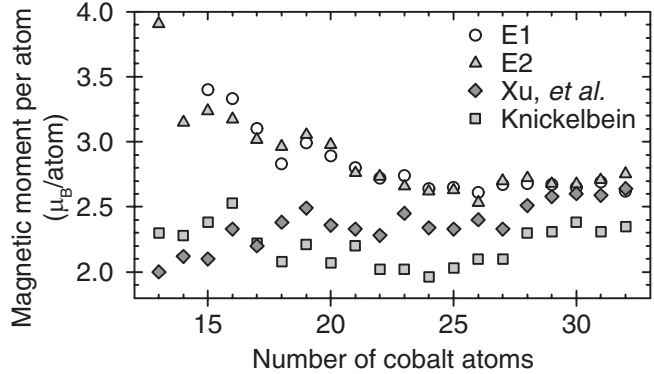}
\caption{\label{fig:smallmoments} Magnetic moments per atom $\mu$ for cobalt clusters consisting of 13--32 atoms. Points marked by hollow circles and shaded triangles correspond to the results of Experiments E1 and E2, respectively. Points marked by shaded diamonds are the results of Xu {\it et al.} from Ref.~\onlinecite{xu2005} and points marked by shaded squares are the results of Knickelbein from Ref.~\onlinecite{knickelbein2006}.}
\end{figure}

Perhaps the most striking difference between the studies appears in the smallest sizes observed. Figure~\ref{fig:smallmoments} shows values of $\mu$ for clusters consisting of between 13 and 32 atoms for all four experiments. While our experiments E1 and E2 found similar $\mu$ values, they differ significantly from those obtained by Knickelbein and by Xu. Not only do few of the $\mu$ values agree between these three groups, there is also little correlation between relative maxima and minima as a function of cluster size. And while our experiments found $\mu$ decreasing with increasing size up to Co$_{26}$, Knickelbein found $\mu$ approximately constant up to that size and Xu found $\mu$ increasing with increasing size over that range.

While these difference could simply reflect problems with one or more of the experiments, they could also stem from more exciting thermal and isomeric issues. Both Xu and Knickelbein studied clusters smaller and colder than we studied, and thereby observed some interesting behaviors. Knickelbein measured a marked change in the deflection profiles of the smallest clusters at low temperatures. These clusters apparently shift from a superparamagnetic state at higher temperature to a non-superparamagnetic state at lower temperature. This is not surprising because studies of supported clusters have shown that the magnetic anisotropy energy per cluster increases dramatically with decreasing cluster size for clusters of fewer than 25 atoms.\cite{gambardella2003}

Both Xu and Knickelbein have also reported the presence of two distinct populations of magnetic isomers in their beams under certain conditions.~\cite{knickelbein2006,electra2006} Although geometric isomers of cobalt clusters have been predicted and studied theoretically,~\cite{miura1994,guevara1997,andriotis1998,rodriguez2003} the calculated differences in their magnetic moments were not as large as those observed in these two experiments.

We found no evidence of a second, less-magnetic population in either of our experiments, so the isomer populations in our cluster beams may have been different from those in the beams of Knickelbein and Xu. When smaller moment isomers are present in a beam and not resolved from the higher moment isomers, they will reduce the average magnetic moments and lead to lower reported values of $\mu$.  If unresolved low-moment isomers are present in the experiments of Knickelbein and Xu but absent in our experiments, that could explain why the values for $\mu$ that Knickelbein and Xu report are smaller than those we measure.

For cobalt clusters with up to 26 atoms, theoretical studies~\cite{guevara1997,andriotis1998,rodriguez2003} have predicted that $\mu$ should decrease with increasing cluster size, consistent with the trend we observe in E1 and E2. One of those theoretical studies~\cite{rodriguez2003} also predicts a sharp minimum in $\mu$ for Co$_{26}$, a feature that appears in both of our experimental results. However, the average $\mu$ values predicted by these calculations are lower than our average $\mu$ value and closer to those of Xu and Knickelbein. Nonetheless, the models used in those calculations don't consider unquenched orbital magnetism, and recent work has shown that electron orbital motion can contribute substantially to the overall magnetic moments of small clusters.~\cite{guirado2003}

For clusters consisting of between 26 and 32 atoms, the $\mu$ values reported by Xu are similar to our $\mu$ values, while those obtained by Knickelbein are still significantly lower. Knickelbein didn't study clusters consisting of more than 32 atoms. All three of the experiments that continue beyond Co$_{32}$ observe a local maximum in $\mu$ at Co$_{36}$--Co$_{37}$, followed by a rapid decrease in $\mu$ as the next 4 or 5 atoms are added. That sudden decline in $\mu$ suggests that a basic structural change---probably a rearrangement of the underlying geometrical lattice---is occurring in the cobalt clusters over this size range. Yang {\it et al.}~\cite{yang1990} and Parks {\it et al.}\cite{parks1993} also found evidence for a structural change in this size range in their ionization potential and chemical reactivity studies, respectively. Another study done recently in our laboratory suggests that free chromium clusters undergo a striking geometrical change at a slightly smaller size (Cr$_{32}$ to Cr$_{34}$).\cite{payne2006} While the theoretical studies of cobalt clusters do not predict a decline in $\mu$, they do not consider every cluster size for these larger clusters.~\cite{guevara1997,rodriguez2003}

\begin{figure}
\includegraphics{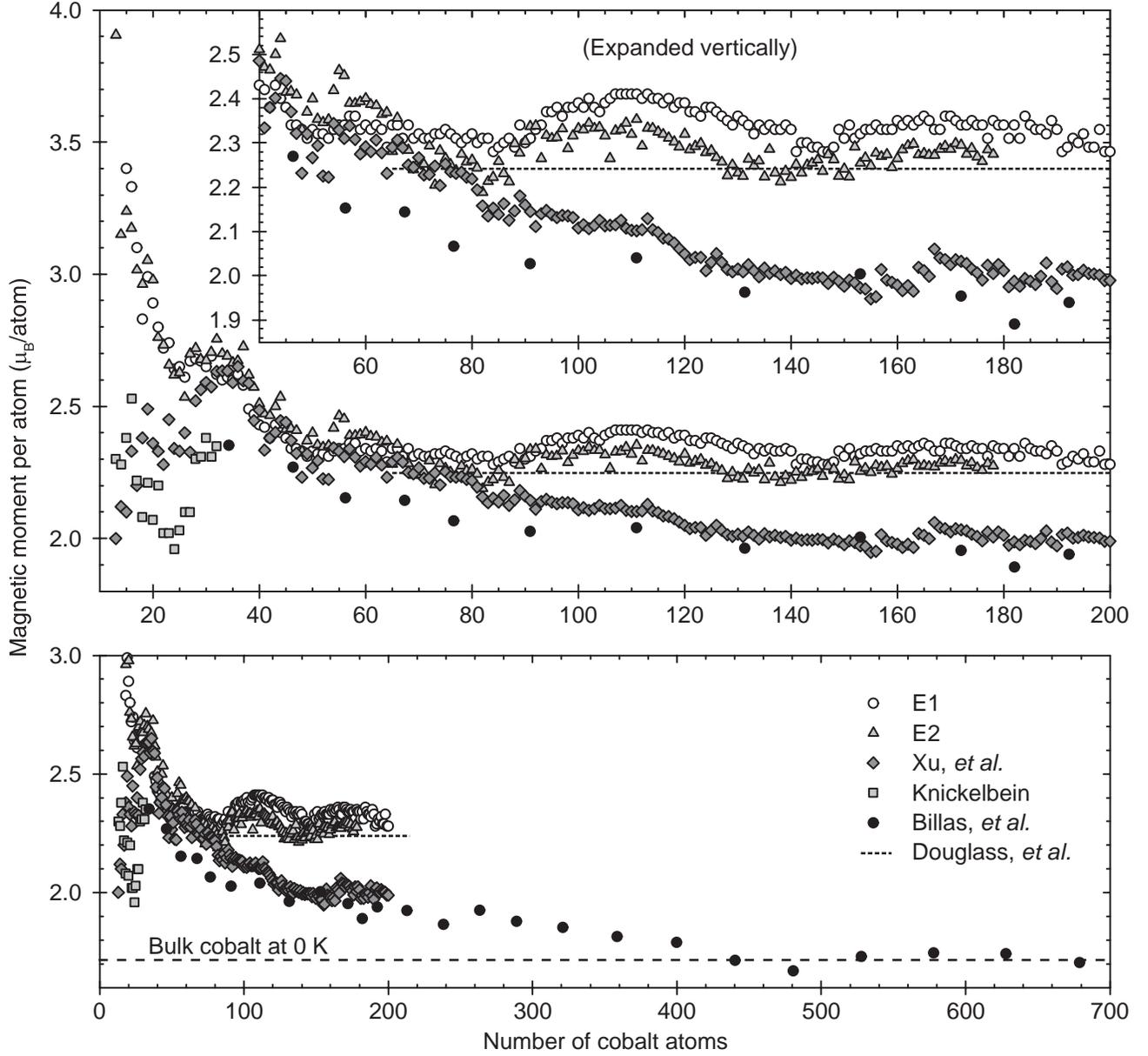}
\caption{\label{fig:allmoments} A review of the measured magnetic moments per atom $\mu$ for cobalt clusters, including both the present work and all of the previous measurements of beam-isolated cobalt clusters. Points marked by hollow circles and shaded triangles correspond to the results of Experiments E1 and E2, respectively. Points marked by shaded diamonds are the results of Xu {\it et al.} from Ref.~\onlinecite{xu2005}, points marked by shaded squares are the results of Knickelbein from Ref.~\onlinecite{knickelbein2006}, and points marked by solid circles are the results of Billas {\it et al.} from Ref.~\onlinecite{billas1994}. The dashed line indicates the single average value measured for Co$_{65-215}$ by Douglass, {\it et al.} (Ref.~\onlinecite{douglass1993}).}
\end{figure}

Above about Co$_{42}$, all three experiments, and the earlier one of Billas, {\it et al.},\cite{billas1994} observed small oscillations in $\mu$ with increasing cluster size (Fig.~\ref{fig:allmoments}). This behavior is typical of clusters of ferromagnetic materials and has been explained in terms of the shell model of cluster growth.~\cite{billas1994,pellarin1994,fujima1996,xie2003} In the shell model, the spectrum of cluster sizes is punctuated occasionally by unusually stable ``closed shells,'' where what has closed is either a geometrical layer of atoms to form complete facets or a set of electronic levels. Long associated with ``magic numbers'' in cluster population distributions, closed-shell clusters are also expected to be local minima for $\mu$ in ferromagnetic clusters. Beyond each shell closing, the value of $\mu$ is expected to drift upward initially before decreasing toward a new minimum as the next larger shell closes. That behavior was proposed more than a decade ago,\cite{billas1994,pellarin1994,fujima1996,xie2003} but extensive calculations on iron clusters have recently bolstered the notion that these variations in $\mu$ are due to geometrical factors.\cite{tiago2006} In their calculations of different lattice structures, Tiago, {\it et al.} showed that faceted arrangements correspond to local minima in magnetic moment per atom $\mu$. 

The differences in the oscillations observed in the three experiments (E1, E2, and Xu) illustrate how difficult it is to make ideal measurements. Not only are there statistical and systematic uncertainties in the measurements, there are also inhomogeneities and irreproducibilities in the cluster samples themselves. In each experiment, the cluster samples must inevitably include minor isomers: clusters with equal numbers of atoms but slight differences in how some of those atoms are arranged on the cluster surfaces. There is now evidence that more significant isomers are also present,~\cite{knickelbein2006,electra2006} isomers with very different magnetic moments and properties and with relative populations that may depend on temperature and/or carrier gas pressure. 

One noticeable difference between our experiments and that of Xu is that our $\mu$ values are consistently about 0.2 $\mu_B$ per atom larger than Xu's. This difference could be due to systematic errors in the measurements themselves. First, Xu's clusters could be hotter than expected, or ours colder. Because each experiment includes data collected at several different temperatures, however, one or both experiments would have to be mistaken about temperature by the same percentage at each temperature studied. 

Second, a systematic error in either lab in the calibration or alignment of the gradient-field magnet, or in the measurement of the beam position in that magnet, could explain the discrepancy. To reduce the possibility of such a systemic error in our lab, we carefully reexamined the beam position and alignment between experiments E1 and E2, using a completely different approach from the one used before E1. It is thus unlikely that errors in our magnet position or alignment are responsible for the discrepancy, although inaccuracy in our magnet calibration remains a possibility. 

There is also an interesting trend to the small differences between the $\mu$ values measured in our two experiments, E1 and E2. For Co$_{30}$ to Co$_{65}$, the measured values of $\mu$ were typically a little lower in E1 than in E2, while for Co$_{65}$ up, the measured values were typically a little higher in E1 than in E2. This switch could be the result of a gas expansion effect in E1. In experiment E1's ``waiting room'' source, gas leaving the source at long residence times may have cooled itself by doing work on gas leaving the source at short residence times. This same cooling effect occurs when you open a can of compressed air.

Since larger clusters take longer to form and cool, we usually study them at longer residence times than the smaller clusters. In E1, those larger clusters may have equilibrated with gas that had cooled itself slightly below the actual temperature of the source and those larger clusters may be slightly colder than we expect. Overestimating their temperatures when fitting data from these clusters to Eq.~(\ref{eq:langevin}) would lead us to overestimate their magnetic moments per atom. In E2, the helium's ongoing thermal equilibration with the long narrow channel through which it flowed should have prevented this sort of excess cooling.

We looked for a cooling effect in the E1 residence time studies themselves, but could not prove its presence or absence convincingly. While the residence time study in Fig.~\ref{fig:restime}(a) may include a slight peak in $\mu$ near 1400~$\mu$s, consistent with a gas temperature that has dropped slightly below the source temperature, the experimental uncertainty is too large for any definitive observations. Still, it would not take much excess cooling in E1 to explain the small differences in $\mu$ between E1 and E2 for Co$_{65}$ up. It is worth noting that almost all of the $\mu$ values obtained in E1 and E2 for Co$_{65}$ up fall within the 2.24$\pm$0.14~$\mu_B$/atom range reported a decade ago in Ref.~\onlinecite{douglass1993} for Co$_{65}$ through Co$_{215}$.

We have interpreted the data from experiments E1 and E2 using Eq.~(\ref{eq:langevin}), which we obtained from the superparamagnetic model. However, the applicability of that model to small, isolated clusters remains a matter of some controversy. In its simplest form, the superparamagnetic model predicts that clusters of a given size should deflect homogeneously to strong field so that there should be no broadening of the deflected beam profile. After observing substantial broadening in the deflected profiles, Xu {\it et al.} proposed an alternative model for the magnetic response of the clusters. In their adiabatic magnetization model, clusters entering a magnetic field shift adiabatically into magnetic states through avoided crossings between Zeeman levels.~\cite{visuthikraisee1996,hamamoto2000,xu2005}

According to this adiabatic magnetization model, clusters deflect towards strong field on average, but diversity of quantum states in the ensemble of clusters entering the magnetic field causes the deflected beam profile to broaden significantly. Despite that broadening, Xu {\it et al.} report that the average magnetization follows the Langevin function and Eq.~(\ref{eq:langevin}),~\cite{xu2005} so that the adiabatic model produces average deflections that are indistinguishable from those produced by the superparamagnetic model.

Xu {\it et al.} contend that the presence of broadening in the deflected beam profile is inconsistent with the superparamagnetic model because the only explanation for such broadening in superparamagnetic clusters is imperfect time averaging---the clusters must fail to explore the entire Boltzmann distribution during the measurement time in the magnet, $t_{\rm mag}$. When the thermal relaxation time $\tau$ is short compared to $t_{\rm mag}$, time averaging is complete and superparamagnetic theory predicts that equal clusters will exhibit equal magnetizations. A beam of such clusters will deflect without broadening. But as $\tau$ approaches $t_{\rm mag}$, time averaging becomes imperfect and equal clusters will begin to deflect unequally.

For incomplete time averaging to be responsible for the broadening observed in the deflection profiles, $\tau$ must be both remarkably long and fortuitously close to $t_{\rm mag}$. Superparamagnetic theory predicts that the width of the magnetic distribution $\Delta M/\mu$ should be approximately equal to $(\tau/t_{\rm mag})^{1/2}$. Xu {\it et al.} found $\Delta M/\mu$ to be insensitive to $t_{\rm mag}$, as well as to T and to N, and therefore rejected superparamagnetism as the explanation for the Langevin-like magnetization behavior.

However, imperfect time averaging is not the only explanation for profile broadening in superparamagnetic clusters. In fact, imperfect time average is the least likely cause of such broadening. $\tau$ is almost certainly in the subnanosecond range,\cite{neel1949,brown1959,khanna1991} so that $(\tau/t_{\rm mag})^{1/2} \approx 10^{-5}$ for most cluster beam deflection experiments. Superparamagnetic clusters thus explore the Boltzmann distribution so quickly compared to the measurement time that time averaging is essentially perfect and equal clusters exhibit equal magnetization. That equality, however, does not rule out broadening caused by the spatial structure of the deflecting magnetic field or by true inequalities between clusters. 

The spatial structure of the deflecting magnetic field is clearly responsible for some of the broadening. Equation~(\ref{eq:langevin}) shows that $\mu_{\rm eff}$, the time-average projection of the cluster's true magnetic moment onto the magnetic field, is approximately proportional to the magnitude of that magnetic field. Since the cluster beam has a finite width and the magnet through which it passes has a gradient magnetic field, clusters on the low field side of the beam will have smaller $\mu_{\rm eff}$ and deflect less than clusters on the high field side. The magnetic field gradient also varies along the height of the cluster beam, perpendicular to the plane of the experiment, further broadening the experimental deflection profiles. 

The second and most important cause of broadening are the inequalities between clusters resulting from the thermal statistics of small systems. The $\mu_{\rm eff}$ given by Eq.~(\ref{eq:langevin}) depends on temperature, but what is the temperature of a particle with only a few dozen atoms? This question is especially appropriate in light of the fact that isolated clusters have several different temperatures (translational, vibrational, and rotational) that are often not in thermal equilibrium with one another.~\cite{bucher1993} 

Thermodynamics is based on the fact that at a temperature, $T$, a system's extensive properties will fluctuate with a variance on the order of $1/N$, where $N$ is the number of particles in the system. It was designed to describe phenomena in macroscopic systems, in which $N$ is very large and the energies associated with those phenomena are much greater than $k_BT$. In nearly infinite systems, fluctuations produce insignificant and essentially undetectable fluctuations in the extensive properties.

But fluctuations are not insignificant in finite systems such as clusters. Recent work has shown that if the energies associated with phenomena in finite systems are of the order $k_BT$, then the fluctuations in the value of an extensive property can be on the order of that property's average value.~\cite{felix2003,bustamante2005,garnier2005} 

In our experiment, the clusters' magnetic energies $N\mu B$ are rarely larger than $k_BT$. For example, for Co$_{100}$ at 100~K and $B=0.15$~T, values typical of this experiment, $N\mu B \simeq 0.2 k_BT $. Fluctuations in observed magnetism should therefore be substantial and could be on the order of those experimental moments themselves. Such statistical distributions in experimental magnetic moments, together with the likely presence of minor isomers, should cause the width of a beam of superparamagnetic clusters to increase significantly as it deflects.

\begin{figure}
\includegraphics{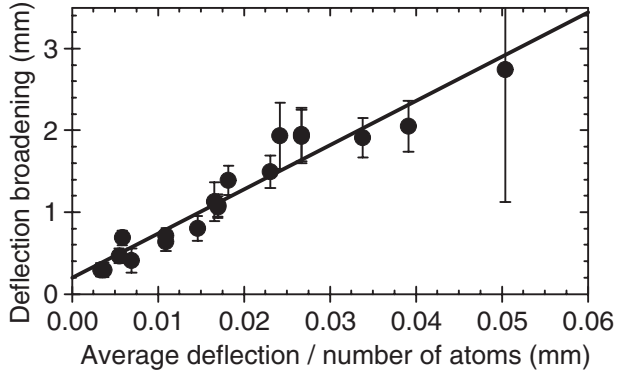}
\caption{\label{fig:fwhm} As cobalt clusters consisting of $N$ atoms deflect toward strong field in our apparatus, their deflection profile broadens and its full width at half maximum increases. Plotted here are experimentally measured broadenings in profile width (i.e., the difference between field-on width and field-off width) as a function of average field-on deflection divided by $N$. The line is a least-squares fit to the experimental points and has an approximate slope of 54.}
\end{figure}

We studied deflection broadening as a function of all the experimental parameters and found empirically that $\Delta d$, the broadening of a deflection profile's full width at half maximum, is proportion to $\bar d$, that cluster's average deflection, divided by the number of atoms in the cluster (Fig.~\ref{fig:fwhm}). That relationship can be written as
\begin{equation}
{{\Delta d}\over{\bar d}} = {C\over N},\label{eq:dspread}
\end{equation}
where $C$ is a constant for cobalt clusters. The slope of the line in Fig.~\ref{fig:fwhm} yields $C\simeq 54$.

In Stern-Gerlach cluster deflection experiments, the experimental moment per atom $\mu_{\rm exp}$ is related to the deflection $d$ according to
\begin{equation}
\mu_{\rm exp} = {{m d v^2}\over {\nabla B(L D + L^2/2)}},\label{eq:deflection}
\end{equation}
where $m$ is the mass per atom, $v$ is the cluster beam velocity, $\nabla B$ is the field gradient, $L$ is the magnet length, and $D$ is the drift region length. Substituting Eq.~(\ref{eq:deflection}) into Eq.~(\ref{eq:dspread}) gives
\begin{equation}
{{\Delta \mu_{\rm exp}}\over {\bar \mu_{\rm exp}}} = {C\over N}.\label{eq:muspread}
\end{equation}
Far from saturation, Eq.~(\ref{eq:langevin}) can be approximated as
\begin{equation}
\mu_{\rm eff} \simeq {{N \mu^2 B}\over {3 k_B T_{\rm vib}}}.\label{eq:langapprox}
\end{equation}
Assuming that $\mu_{\rm eff}$ in Eq.~(\ref{eq:langapprox}) is $\mu_{\rm exp}$, Eq.~(\ref{eq:muspread}) becomes
\begin{equation}
{{\Delta T_{\rm vib}}\over {\bar T_{\rm vib}}} = {C\over N},\label{eq:Tspread}
\end{equation}
where $\Delta T_{\rm vib}$ is defined as a positive quantity in order to eliminate a meaningless negative sign.

This analysis suggests that the clusters have a fractional spread in vibrational temperature $\Delta T_{\rm vib}/{\bar T_{\rm vib}}$ that is proportional to $C/N$. That this fractional spread increases with decreasing cluster size as $1/N$ is consistent with the statistics of finite systems and the inevitable deviations from the equipartition theorem.

One hallmark of superparamagnetic behavior is that the clusters should never deflect toward weak field, even when broadening effects are taken into account. The adiabatic magnetization model, on the other hand, does not exclude weak-field deflection. We looked carefully for deflection to weak field but could not find it in any of our measurements.

In fact, when we studied the deflection of the cluster beam's low-field edge by strong magnetic fields and field gradients, we found that we could deflect that edge of the beam toward strong-field by as much as 1~mm. This deflection was complete in that there were no clusters left undeflected at the beam's low-field edge. If any of the clusters had deflected toward weak-field or avoided deflection altogether, there would not have been an empty region along the low-field side of the beam. We found no clusters there, indicating that all of the clusters deflected toward strong-field, consistent with superparamagnetism.

Although both cluster magnetism models, superparamagnetism and adiabatic magnetization, yield Langevin-like relationships between the effective moment per atom $\mu_{\rm eff}$ and the true moment per atom $\mu$, the temperature that is relevant to the Langevin function is different for the two models. Clusters in a seeded supersonic beam have at least three different temperatures, a translational temperature $T_{\rm trans}$, a rotational temperature $T_{\rm rot}$, and a vibrational temperature $T_{\rm vib}$, and those three temperatures are unlikely to be in equilibrium with one another.

It is well established that when molecules or clusters are seeded in a noble gas and undergo supersonic or free jet expansion, their translational and rotational temperatures cool far more easily than their vibrational temperatures.\cite{amirav1980,geusic1985,wallraff1987,douglass1993} In most cases, $T_{\rm trans} < T_{\rm rot} < T_{\rm vib}$. For example, Wallraff {\it et al.} measured $T_{\rm rot} = 5.7$~K and $T_{\rm vib} = 213$~K in a beam of the triatomic molecule OCS from a free jet expansion in room-temperature argon.\cite{wallraff1987} While it is relatively straightforward to produce a supersonic beam of clusters with $T_{\rm vib} \simeq T_{\rm source}$,\cite{collings1993} it is much harder to achieve $T_{\rm rot} \simeq T_{\rm source}$. Rotational cooling is so efficient in seeded supersonic expansions that clusters in the resulting beam will almost certainly have $T_{\rm rot} << T_{\rm source}$.

A second important thermal issue is the heat capacity associated with each cluster temperature. Xu {\it et al.} assert that a cluster's vibrational degrees of freedom should freeze out when $T_{\rm vib} < T_{\rm Debye}/N^{1/3}$ and therefore that most of the smaller clusters they study are in their ground vibrational states.\cite{xu2005} This suggestion that the clusters effectively have zero vibrational heat capacity at the experimental temperatures neglects the softening that occurs when a solid is reduced to cluster dimensions. With most of its atoms on its surface, a cluster is no longer bulklike with respect to phonons and vibrations,\cite{frenken1985,bohr2001} as evidenced by the reduced melting temperatures\cite{bachels2000} and premelting effects\cite{breaux2005} seen in nanoparticles and clusters. Instead, it retains considerable vibrational heat capacity even at relatively low temperatures. For example, carbon has a Debye temperature of 2230~K, yet C$_6$ has vibrational energies as low as $\sim 100$~cm$^{-1}$ ($\sim140$~K).\cite{zhao1996,xu1997}

In superparamagnetism, each cluster acts as its own heat bath and that heat bath is dominated by the cluster's vast number of vibrational degrees of freedom. Moreover, the cluster's three rotational degrees of freedom are highly constrained by angular momentum conservation. As a result, the effective temperature of the cluster's internal heat bath and the temperature that must appear in the Langevin function is $T_{\rm vib}$. Fortunately, $T_{\rm vib}$ is relatively easy to control in a seeded supersonic cluster beam.

In the adiabatic magnetization model, however, the assumption that spin-rotation coupling is responsible for the Langevin-like behavior implies that the rotational temperature $T_{\rm rot}$ is dominant. Moreover, if the vibrations are frozen out, $T_{\rm vib}$ has no role at all in this model. The temperature that should appear in the Langevin function is $T_{\rm rot}$. Since $T_{\rm rot}$ is difficult to control, and unlikely to be even close to the $T_{\rm source}$, measurements of magnetic properties that depend on $T_{\rm rot}$ will be difficult to interpret.

Each cluster has a geometrical lattice and the cluster's magnetic moment tends to align along an easy axis of that lattice. The energy required to reorient the moment away from an easy lattice axis is known as the magnetocrystalline anisotropy energy (MAE). A familiar concept for bulk materials, MAEs have also been calculated for cobalt clusters and are found to depend strongly on the size and detailed structure of a cluster.\cite{guirado2001,xie2004} For cobalt clusters of the sizes we studied, the MAEs are predicted to range from approximately 30 $\mu$eV (0.4 K) for highly symmetric and/or small clusters to 400 $\mu$eV (5 K) for less symmetric and/or large clusters. Since $T_{\rm vib} \ge 60$~K for all of our measurements, $k_BT$ is large compared to the MAE and our clusters should be well into the superparamagnetic regime.

For cobalt clusters to avoid superparamagnetic fluctuations, they would have to be cooled to extremely low temperatures or have anomalously large MAEs. Knickelbein reports observing non-superparamagnetic or ``locked-moment behavior''\cite{douglass1993} in Co$_{7-11}$ at temperatures below 100 K and in Co$_{15}$ at temperatures near 50 K.\cite{knickelbein2006}

MAEs also have a role in the adiabatic magnetization model proposed by Xu, {\it et al.} in Ref.~\onlinecite{xu2005}. In that model, the Zeeman levels arising from a cluster's magnetic spin quantum number $S$ fan out in energy with increasing magnetic field. Since each rotational level associated with the cluster's rotational quantum number $J$ has its own manifold of Zeeman sublevels, the Zeeman fans of different rotational levels encounter one another as the field increases. Couplings between spin and rotation cause these Zeeman levels to experience avoided crossing. The adiabatic magnetization model proposes that as each cluster enters a magnetic field, its state evolves adiabatically through the avoided crossings in such a way that an ensemble of clusters exhibits an average magnetization that depends on the Langevin function Eq~(\ref{eq:langevin}).

The spin-rotation coupling responsible for these avoided crossings is the MAE and without that coupling, the Zeeman levels would simply cross one another. But this Zeeman-level picture assumes that spin-rotation coupling is relatively weak compared to the level spacings so that a cluster's spin angular momentum $S$ and rotational angular momentum $J$ are both good quantum numbers.

In most of the experimental conditions, however, the rotational energy spacings are less than the MAE. For example, at 60 K, Co$_{55}$ has $J \approx 270$ and the energy spacing between $J$ and $J+1$ is approximately 30 $\mu$eV. Since the predicted MAE for Co$_{55}$ is also approximately 30 $\mu$eV, the coupling energy is comparable to the spacings between rotational levels and the Zeeman picture blurs into a sea of mixed states. For larger or less-symmetric cobalt clusters, the spin-rotation couplings are even larger. A detailed theory of these spin-rotation effects should therefore be rather complicated.

\section{Conclusion}
\label{Conclusion}
We have measured the magnetic moments per atom for small cobalt clusters (13--200 atoms) and find them size dependent. All of these clusters have magnetic moments per atom that are larger than the value for bulk cobalt (1.7 $\mu_B$/atom), evidence that the reduced dimensionality and increased surface to volume ratio of clusters leads to enhanced magnetism. We see gradual magnetic oscillations with increasing size, consistent with a shell model of cluster growth. Changes made to our experimental apparatus and data acquisition procedure were found to have had little impact on the results. We find that superparamagnetism and the statistics of finite systems can explain both our experimentally observed magnetic moments per atom and the spreads in the deflection profiles of all of the clusters. We have compared the superparamagnetic and adiabatic magnetization models with the experimental and theoretical evidence and find that the superparamagnetic model is the most likely explanation for the Langevin-like magnetizations of the isolated cobalt clusters in our experiments. 

\begin{acknowledgments}
One of the authors (LAB) acknowledges useful conversations with Thomas F. Gallagher. This material is based upon work supported by the National Science Foundation under Grant No. DMR-0405203.
\end{acknowledgments}

\begin{longtable}{|c|c|c||c|c|c||c|c|c||c|c|c|}
\caption[Magnetic moments per atom, $\mu$, obtained by applying Eq. (\ref{eq:langevin}) to the experimentally measured moments per atom, $\mu_{\rm eff}$. Only statistical uncertainties are indicated (1~SD).]{Magnetic moments per atom, $\mu$, obtained by applying Eq. (\ref{eq:langevin}) to the experimentally measured moments per atom, $\mu_{\rm eff}$. Only statistical uncertainties are indicated (1~SD).} \label{tab:moments} \\

\hline
& \multicolumn{2}{|c||} {$\mu$ ($\mu_B$)} & 
& \multicolumn{2}{|c||} {$\mu$ ($\mu_B$)} & 
& \multicolumn{2}{|c||} {$\mu$ ($\mu_B$)} & 
& \multicolumn{2}{|c|} {$\mu$ ($\mu_B$)}\\
N&E1&E2&
N&E1&E2&
N&E1&E2&
N&E1&E2\\
\hline
\endfirsthead

\hline
& \multicolumn{2}{|c||} {$\mu$ ($\mu_B$)} & 
& \multicolumn{2}{|c||} {$\mu$ ($\mu_B$)} & 
& \multicolumn{2}{|c||} {$\mu$ ($\mu_B$)} & 
& \multicolumn{2}{|c|} {$\mu$ ($\mu_B$)}\\
N&E1&E2&
N&E1&E2&
N&E1&E2&
N&E1&E2\\
\hline
\endhead

13 & & 3.91$\pm$.60 & 60 & 2.34$\pm$.01 & 2.40$\pm$.02 & 107 & 
2.41$\pm$.01 & 2.33$\pm$.02 & 154 & 2.33$\pm$.02 & 2.27$\pm$.02 \\
14 & & 3.15$\pm$.33 & 61 & 2.33$\pm$.02 & 2.39$\pm$.02 & 108 & 
2.41$\pm$.01 & 2.33$\pm$.02 & 155 & 2.33$\pm$.02 & 2.25$\pm$.02 \\
15 & 3.40$\pm$.16 & 3.24$\pm$.17 & 62 & 2.34$\pm$.02 & 2.38$\pm$.02 
& 109 & 2.41$\pm$.01 & 2.34$\pm$.02 & 156 & 0.34$\pm$.02 & 
2.28$\pm$.02 \\
16 & 3.33$\pm$.23 & 3.17$\pm$.23 & 63 & 2.33$\pm$.01 & 2.36$\pm$.02 
& 110 & 2.41$\pm$.01 & 2.32$\pm$.02 & 157 & 2.33$\pm$.02 & 
2.27$\pm$.02 \\
17 & 3.10$\pm$.10 & 3.02$\pm$.09 & 64 & 2.29$\pm$.02 & 2.37$\pm$.02 
& 111 & 2.41$\pm$.01 & 2.35$\pm$.02 & 158 & 2.33$\pm$.02 & 
2.27$\pm$.02 \\
18 & 2.83$\pm$.08 & 2.96$\pm$.12 & 65 & 2.34$\pm$.01 & 2.28$\pm$.02 
& 112 & 2.40$\pm$.01 & 2.29$\pm$.02 & 159 & 2.33$\pm$.01 & 
2.24$\pm$.02 \\
19 & 2.99$\pm$.04 & 3.05$\pm$.07 & 66 & 2.34$\pm$.01 & 2.36$\pm$.02 
& 113 & 2.41$\pm$.01 & 2.33$\pm$.02 & 160 & 2.35$\pm$.02 & 
2.26$\pm$.02 \\
20 & 2.89$\pm$.06 & 2.98$\pm$.06 & 67 & 2.31$\pm$.01 & 2.29$\pm$.02 
& 114 & 2.40$\pm$.01 & 2.33$\pm$.02 & 161 & 2.34$\pm$.02 & 
2.28$\pm$.02 \\
21 & 2.80$\pm$.05 & 2.76$\pm$.07 & 68 & 2.34$\pm$.01 & 2.28$\pm$.02 
& 115 & 2.40$\pm$.01 & 2.33$\pm$.02 & 162 & 2.35$\pm$.02 & 
2.29$\pm$.02 \\
22 & 2.72$\pm$.05 & 2.73$\pm$.06 & 69 & 2.31$\pm$.01 & 2.24$\pm$.02 
& 116 & 2.39$\pm$.01 & 2.32$\pm$.02 & 163 & 2.34$\pm$.02 & 
2.30$\pm$.02 \\
23 & 2.74$\pm$.04 & 2.66$\pm$.04 & 70 & 2.32$\pm$.01 & 2.26$\pm$.02 
& 117 & 2.40$\pm$.01 & 2.31$\pm$.02 & 164 & 2.35$\pm$.02 & 
2.27$\pm$.02 \\
24 & 2.64$\pm$.03 & 2.62$\pm$.05 & 71 & 2.31$\pm$.01 & 2.25$\pm$.02 
& 118 & 2.38$\pm$.01 & 2.30$\pm$.02 & 165 & 2.36$\pm$.02 & 
2.28$\pm$.02 \\
25 & 2.65$\pm$.04 & 2.63$\pm$.04 & 72 & 2.31$\pm$.00 & 2.29$\pm$.02 
& 119 & 2.39$\pm$.01 & 2.29$\pm$.02 & 166 & 2.35$\pm$.02 & 
2.27$\pm$.02 \\
26 & 2.61$\pm$.03 & 2.53$\pm$.06 & 73 & 2.32$\pm$.01 & 2.20$\pm$.02 
& 120 & 2.39$\pm$.01 & 2.32$\pm$.02 & 167 & 2.33$\pm$.02 & 
2.27$\pm$.02 \\
27 & 2.67$\pm$.02 & 2.70$\pm$.04 & 74 & 2.32$\pm$.01 & 2.28$\pm$.02 
& 121 & 2.37$\pm$.01 & 2.28$\pm$.02 & 168 & 2.33$\pm$.02 & 
2.29$\pm$.02 \\
28 & 2.68$\pm$.03 & 2.72$\pm$.04 & 75 & 2.33$\pm$.01 & 2.26$\pm$.02 
& 122 & 2.37$\pm$.01 & 2.29$\pm$.02 & 169 & 2.36$\pm$.02 & 
2.29$\pm$.02 \\
29 & 2.67$\pm$.03 & 2.68$\pm$.03 & 76 & 2.32$\pm$.01 & 2.25$\pm$.02 
& 123 & 2.36$\pm$.01 & 2.30$\pm$.02 & 170 & 2.36$\pm$.02 & 
2.29$\pm$.02 \\
30 & 2.65$\pm$.04 & 2.67$\pm$.04 & 77 & 2.31$\pm$.01 & 2.24$\pm$.02 
& 124 & 2.38$\pm$.01 & 2.28$\pm$.02 & 171 & 2.35$\pm$.02 & 
2.30$\pm$.02 \\
31 & 2.69$\pm$.03 & 2.71$\pm$.03 & 78 & 2.30$\pm$.01 & 2.27$\pm$.02 
& 125 & 2.38$\pm$.01 & 2.28$\pm$.02 & 172 & 2.34$\pm$.01 & 
2.29$\pm$.02 \\
32 & 2.62$\pm$.03 & 2.75$\pm$.03 & 79 & 2.31$\pm$.01 & 2.26$\pm$.02 
& 126 & 2.37$\pm$.01 & 2.27$\pm$.02 & 173 & 2.34$\pm$.02 & 
2.27$\pm$.02 \\
33 & 2.60$\pm$.03 & 2.70$\pm$.03 & 80 & 2.32$\pm$.01 & 2.26$\pm$.02 
& 127 & 2.36$\pm$.01 & 2.26$\pm$.03 & 174 & 2.35$\pm$.02 & 
2.29$\pm$.02 \\
34 & 2.61$\pm$.02 & 2.69$\pm$.03 & 81 & 2.30$\pm$.01 & 2.24$\pm$.02 
& 128 & 2.35$\pm$.01 & 2.23$\pm$.03 & 175 & 2.34$\pm$.02 & 
2.27$\pm$.02 \\
35 & 2.60$\pm$.02 & 2.66$\pm$.02 & 82 & 2.31$\pm$.01 & 2.19$\pm$.02 
& 129 & 2.34$\pm$.01 & 2.25$\pm$.03 & 176 & 2.34$\pm$.01 & 
2.26$\pm$.03 \\
36 & 2.62$\pm$.02 & 2.67$\pm$.03 & 83 & 2.31$\pm$.01 & 2.21$\pm$.02 
& 130 & 2.36$\pm$.01 & 2.23$\pm$.03 & 177 & 2.31$\pm$.02 & 
2.28$\pm$.03 \\
37 & 2.58$\pm$.02 & 2.73$\pm$.02 & 84 & 2.28$\pm$.01 & 2.22$\pm$.02 
& 131 & 2.35$\pm$.01 & 2.22$\pm$.03 & 178 & 2.35$\pm$.02 & 
2.27$\pm$.05 \\
38 & 2.49$\pm$.02 & 2.62$\pm$.02 & 85 & 2.29$\pm$.01 & 2.26$\pm$.02 
& 132 & 2.34$\pm$.01 & 2.26$\pm$.02 & 179 & 2.34$\pm$.02 & \\
39 & 2.47$\pm$.02 & 2.57$\pm$.02 & 86 & 2.30$\pm$.01 & 2.23$\pm$.02 
& 133 & 2.32$\pm$.01 & 2.25$\pm$.03 & 180 & 2.34$\pm$.02 & \\
40 & 2.43$\pm$.02 & 2.51$\pm$.02 & 87 & 2.31$\pm$.01 & 2.21$\pm$.02 
& 134 & 2.34$\pm$.01 & 2.26$\pm$.02 & 181 & 2.31$\pm$.01 & \\
41 & 2.42$\pm$.02 & 2.47$\pm$.02 & 88 & 2.29$\pm$.01 & 2.30$\pm$.03 
& 135 & 2.33$\pm$.01 & 2.22$\pm$.03 & 182 & 2.34$\pm$.02 & \\
42 & 2.38$\pm$.02 & 2.46$\pm$.02 & 89 & 2.34$\pm$.01 & 2.28$\pm$.03 
& 136 & 2.34$\pm$.01 & 2.29$\pm$.03 & 183 & 2.31$\pm$.01 & \\
43 & 2.43$\pm$.02 & 2.50$\pm$.02 & 90 & 2.30$\pm$.01 & 2.30$\pm$.03 
& 137 & 2.33$\pm$.01 & 2.23$\pm$.03 & 184 & 2.36$\pm$.02 & \\
44 & 2.40$\pm$.02 & 2.54$\pm$.02 & 91 & 2.31$\pm$.01 & 2.34$\pm$.03 
& 138 & 2.33$\pm$.01 & 2.21$\pm$.03 & 185 & 2.34$\pm$.02 & \\
45 & 2.38$\pm$.02 & 2.41$\pm$.02 & 92 & 2.33$\pm$.01 & 2.34$\pm$.03 
& 139 & 2.34$\pm$.01 & 2.24$\pm$.02 & 186 & 2.33$\pm$.02 & \\
46 & 2.34$\pm$.02 & 2.42$\pm$.02 & 93 & 2.34$\pm$.01 & 2.26$\pm$.03 
& 140 & 2.33$\pm$.01 & 2.22$\pm$.02 & 187 & 2.33$\pm$.02 & \\
47 & 2.34$\pm$.02 & 2.41$\pm$.02 & 94 & 2.37$\pm$.01 & 2.32$\pm$.02 
& 141 & 2.28$\pm$.01 & 2.25$\pm$.02 & 188 & 2.32$\pm$.02 & \\
48 & 2.33$\pm$.02 & 2.33$\pm$.02 & 95 & 2.37$\pm$.01 & 2.31$\pm$.02 
& 142 & 2.30$\pm$.01 & 2.23$\pm$.02 & 189 & 2.35$\pm$.01 & \\
49 & 2.31$\pm$.02 & 2.37$\pm$.02 & 96 & 2.38$\pm$.01 & 2.31$\pm$.03 
& 143 & 2.30$\pm$.01 & 2.25$\pm$.03 & 190 & 2.33$\pm$.02 & \\
50 & 2.33$\pm$.02 & 2.40$\pm$.02 & 97 & 2.36$\pm$.01 & 2.33$\pm$.02 
& 144 & 2.29$\pm$.01 & 2.25$\pm$.02 & 191 & 2.28$\pm$.02 & \\
51 & 2.32$\pm$.02 & 2.35$\pm$.01 & 98 & 2.38$\pm$.01 & 2.29$\pm$.02 
& 145 & 2.29$\pm$.02 & 2.26$\pm$.02 & 192 & 2.29$\pm$.03 & \\
52 & 2.32$\pm$.02 & 2.35$\pm$.02 & 99 & 2.38$\pm$.01 & 2.32$\pm$.02 
& 146 & 2.28$\pm$.02 & 2.24$\pm$.03 & 193 & 2.31$\pm$.02 & \\
53 & 2.31$\pm$.02 & 2.35$\pm$.02 & 100 & 2.39$\pm$.01 & 
2.33$\pm$.02 & 147 & 2.28$\pm$.02 & 2.27$\pm$.02 & 194 & 
2.30$\pm$.02 & \\
54 & 2.33$\pm$.02 & 2.42$\pm$.02 & 101 & 2.38$\pm$.01 & 
2.33$\pm$.02 & 148 & 2.29$\pm$.01 & 2.26$\pm$.02 & 195 & 
2.32$\pm$.03 & \\
55 & 2.34$\pm$.02 & 2.46$\pm$.02 & 102 & 2.40$\pm$.01 & 
2.34$\pm$.02 & 149 & 2.33$\pm$.02 & 2.22$\pm$.02 & 196 & 
2.29$\pm$.03 & \\
56 & 2.34$\pm$.02 & 2.45$\pm$.01 & 103 & 2.38$\pm$.01 & 
2.33$\pm$.02 & 150 & 2.31$\pm$.02 & 2.24$\pm$.02 & 197 & 
2.29$\pm$.03 & \\
57 & 2.36$\pm$.02 & 2.39$\pm$.02 & 104 & 2.37$\pm$.01 & 
2.33$\pm$.02 & 151 & 2.34$\pm$.02 & 2.22$\pm$.02 & 198 & 
2.33$\pm$.03 & \\
58 & 2.36$\pm$.02 & 2.39$\pm$.02 & 105 & 2.39$\pm$.01 & 
2.32$\pm$.02 & 152 & 2.31$\pm$.02 & 2.29$\pm$.02 & 199 & 
2.28$\pm$.02 & \\
59 & 2.33$\pm$.02 & 2.39$\pm$.02 & 106 & 2.40$\pm$.01 & 
2.26$\pm$.02 & 153 & 2.32$\pm$.01 & 2.25$\pm$.02 & 200 & 
2.28$\pm$.02 & \\
\hline
\end{longtable}

\bibliography{cobalt}

\begin{thebibliography}{58}
\expandafter\ifx\csname natexlab\endcsname\relax\def\natexlab#1{#1}\fi
\expandafter\ifx\csname bibnamefont\endcsname\relax
  \def\bibnamefont#1{#1}\fi
\expandafter\ifx\csname bibfnamefont\endcsname\relax
  \def\bibfnamefont#1{#1}\fi
\expandafter\ifx\csname citenamefont\endcsname\relax
  \def\citenamefont#1{#1}\fi
\expandafter\ifx\csname url\endcsname\relax
  \def\url#1{\texttt{#1}}\fi
\expandafter\ifx\csname urlprefix\endcsname\relax\def\urlprefix{URL }\fi
\providecommand{\bibinfo}[2]{#2}
\providecommand{\eprint}[2][]{\url{#2}}

\bibitem[{\citenamefont{Neel}(1949)}]{neel1949}
\bibinfo{author}{\bibfnamefont{L.}~\bibnamefont{Neel}},
  \bibinfo{journal}{Compt. Rend.} \textbf{\bibinfo{volume}{228}},
  \bibinfo{pages}{664} (\bibinfo{year}{1949}).

\bibitem[{\citenamefont{Bean}(1955)}]{bean1955}
\bibinfo{author}{\bibfnamefont{C.~P.} \bibnamefont{Bean}}, \bibinfo{journal}{J
  Appl. Phys.} \textbf{\bibinfo{volume}{26}}, \bibinfo{pages}{1381}
  (\bibinfo{year}{1955}).

\bibitem[{\citenamefont{Jacobs and Bean}(1963)}]{jacobs1963}
\bibinfo{author}{\bibfnamefont{I.~S.} \bibnamefont{Jacobs}} \bibnamefont{and}
  \bibinfo{author}{\bibfnamefont{C.~P.} \bibnamefont{Bean}}, in
  \emph{\bibinfo{booktitle}{Magnetism}}, edited by
  \bibinfo{editor}{\bibfnamefont{G.~T.} \bibnamefont{Rado}} \bibnamefont{and}
  \bibinfo{editor}{\bibfnamefont{H.}~\bibnamefont{Suhl}}
  (\bibinfo{publisher}{Academic}, \bibinfo{year}{1963}), vol.
  \bibinfo{volume}{III}, pp. \bibinfo{pages}{271--350}.

\bibitem[{\citenamefont{Bean and Livingston}(1959)}]{bean1959}
\bibinfo{author}{\bibfnamefont{C.~P.} \bibnamefont{Bean}} \bibnamefont{and}
  \bibinfo{author}{\bibfnamefont{J.~D.} \bibnamefont{Livingston}},
  \bibinfo{journal}{J. Appl. Phys.} \textbf{\bibinfo{volume}{30}},
  \bibinfo{pages}{120S} (\bibinfo{year}{1959}).

\bibitem[{\citenamefont{Brown}(1959)}]{brown1959}
\bibinfo{author}{\bibfnamefont{W.~F.} \bibnamefont{Brown}},
  \bibinfo{journal}{J. Appl. Phys.} \textbf{\bibinfo{volume}{30}},
  \bibinfo{pages}{130S} (\bibinfo{year}{1959}).

\bibitem[{\citenamefont{Liu et~al.}(1991)\citenamefont{Liu, Khanna, and
  Jena}}]{liu1991}
\bibinfo{author}{\bibfnamefont{F.}~\bibnamefont{Liu}},
  \bibinfo{author}{\bibfnamefont{S.~N.} \bibnamefont{Khanna}},
  \bibnamefont{and} \bibinfo{author}{\bibfnamefont{P.}~\bibnamefont{Jena}},
  \bibinfo{journal}{Phys. Rev. B} \textbf{\bibinfo{volume}{43}},
  \bibinfo{pages}{8179} (\bibinfo{year}{1991}).

\bibitem[{\citenamefont{Dunlap}(1991)}]{dunlap1991}
\bibinfo{author}{\bibfnamefont{B.~I.} \bibnamefont{Dunlap}},
  \bibinfo{journal}{Z. Phys. D} \textbf{\bibinfo{volume}{19}},
  \bibinfo{pages}{255} (\bibinfo{year}{1991}).

\bibitem[{\citenamefont{Bucher et~al.}(1991)\citenamefont{Bucher, Douglass, and
  Bloomfield}}]{bucher1991}
\bibinfo{author}{\bibfnamefont{J.~P.} \bibnamefont{Bucher}},
  \bibinfo{author}{\bibfnamefont{D.~C.} \bibnamefont{Douglass}},
  \bibnamefont{and} \bibinfo{author}{\bibfnamefont{L.~A.}
  \bibnamefont{Bloomfield}}, \bibinfo{journal}{Phys. Rev. Lett.}
  \textbf{\bibinfo{volume}{66}}, \bibinfo{pages}{3052} (\bibinfo{year}{1991}).

\bibitem[{\citenamefont{Billas et~al.}(1993)\citenamefont{Billas, Becker,
  Chatelain, and de~Heer}}]{billas1993}
\bibinfo{author}{\bibfnamefont{I.~M.~L.} \bibnamefont{Billas}},
  \bibinfo{author}{\bibfnamefont{J.~A.} \bibnamefont{Becker}},
  \bibinfo{author}{\bibfnamefont{A.}~\bibnamefont{Chatelain}},
  \bibnamefont{and} \bibinfo{author}{\bibfnamefont{W.~A.}
  \bibnamefont{de~Heer}}, \bibinfo{journal}{Phys. Rev. Lett.}
  \textbf{\bibinfo{volume}{71}}, \bibinfo{pages}{4067} (\bibinfo{year}{1993}).

\bibitem[{\citenamefont{Billas et~al.}(1994)\citenamefont{Billas, Chatelain,
  and de~Heer}}]{billas1994}
\bibinfo{author}{\bibfnamefont{I.~M.~L.} \bibnamefont{Billas}},
  \bibinfo{author}{\bibfnamefont{A.}~\bibnamefont{Chatelain}},
  \bibnamefont{and} \bibinfo{author}{\bibfnamefont{W.~A.}
  \bibnamefont{de~Heer}}, \bibinfo{journal}{Science}
  \textbf{\bibinfo{volume}{265}}, \bibinfo{pages}{1682} (\bibinfo{year}{1994}).

\bibitem[{\citenamefont{Apsel et~al.}(1996)\citenamefont{Apsel, Emmert, Deng,
  and Bloomfield}}]{apsel1996}
\bibinfo{author}{\bibfnamefont{S.~E.} \bibnamefont{Apsel}},
  \bibinfo{author}{\bibfnamefont{J.~W.} \bibnamefont{Emmert}},
  \bibinfo{author}{\bibfnamefont{J.}~\bibnamefont{Deng}}, \bibnamefont{and}
  \bibinfo{author}{\bibfnamefont{L.~A.} \bibnamefont{Bloomfield}},
  \bibinfo{journal}{Phys. Rev. Lett.} \textbf{\bibinfo{volume}{76}},
  \bibinfo{pages}{1441} (\bibinfo{year}{1996}).

\bibitem[{\citenamefont{Xu et~al.}(2005)\citenamefont{Xu, Yin, Moro, and
  de~Heer}}]{xu2005}
\bibinfo{author}{\bibfnamefont{X.}~\bibnamefont{Xu}},
  \bibinfo{author}{\bibfnamefont{S.}~\bibnamefont{Yin}},
  \bibinfo{author}{\bibfnamefont{R.}~\bibnamefont{Moro}}, \bibnamefont{and}
  \bibinfo{author}{\bibfnamefont{W.~A.} \bibnamefont{de~Heer}},
  \bibinfo{journal}{Phys. Rev. Lett} \textbf{\bibinfo{volume}{95}},
  \bibinfo{pages}{237209} (\bibinfo{year}{2005}).

\bibitem[{\citenamefont{Cox et~al.}(1993)\citenamefont{Cox, Louderback, and
  Bloomfield}}]{cox1993}
\bibinfo{author}{\bibfnamefont{A.~J.} \bibnamefont{Cox}},
  \bibinfo{author}{\bibfnamefont{J.~G.} \bibnamefont{Louderback}},
  \bibnamefont{and} \bibinfo{author}{\bibfnamefont{L.~A.}
  \bibnamefont{Bloomfield}}, \bibinfo{journal}{Phys. Rev. Lett}
  \textbf{\bibinfo{volume}{71}}, \bibinfo{pages}{923} (\bibinfo{year}{1993}).

\bibitem[{\citenamefont{Knickelbein}(2001)}]{knickelbein2001}
\bibinfo{author}{\bibfnamefont{M.~B.} \bibnamefont{Knickelbein}},
  \bibinfo{journal}{Phys. Rev. Lett.} \textbf{\bibinfo{volume}{86}},
  \bibinfo{pages}{5255} (\bibinfo{year}{2001}).

\bibitem[{\citenamefont{Douglass et~al.}(1992)\citenamefont{Douglass, Bucher,
  and Bloomfield}}]{douglass1992}
\bibinfo{author}{\bibfnamefont{D.~C.} \bibnamefont{Douglass}},
  \bibinfo{author}{\bibfnamefont{J.~P.} \bibnamefont{Bucher}},
  \bibnamefont{and} \bibinfo{author}{\bibfnamefont{L.~A.}
  \bibnamefont{Bloomfield}}, \bibinfo{journal}{Phys. Rev. B}
  \textbf{\bibinfo{volume}{45}}, \bibinfo{pages}{6341} (\bibinfo{year}{1992}).

\bibitem[{\citenamefont{Bucher and Bloomfiled}(1993)}]{bucher1993}
\bibinfo{author}{\bibfnamefont{J.~P.} \bibnamefont{Bucher}} \bibnamefont{and}
  \bibinfo{author}{\bibfnamefont{L.~A.} \bibnamefont{Bloomfiled}},
  \bibinfo{journal}{Int. J. Mod. Phys. B} \textbf{\bibinfo{volume}{7}},
  \bibinfo{pages}{1079} (\bibinfo{year}{1993}).

\bibitem[{\citenamefont{Cox et~al.}(1994)\citenamefont{Cox, Louderback, Apsel,
  and Bloomfield}}]{cox1994}
\bibinfo{author}{\bibfnamefont{A.~J.} \bibnamefont{Cox}},
  \bibinfo{author}{\bibfnamefont{J.~G.} \bibnamefont{Louderback}},
  \bibinfo{author}{\bibfnamefont{S.~E.} \bibnamefont{Apsel}}, \bibnamefont{and}
  \bibinfo{author}{\bibfnamefont{L.~A.} \bibnamefont{Bloomfield}},
  \bibinfo{journal}{Phys. Rev. B} \textbf{\bibinfo{volume}{49}},
  \bibinfo{pages}{12295} (\bibinfo{year}{1994}).

\bibitem[{\citenamefont{Knickelbein}(2006)}]{knickelbein2006}
\bibinfo{author}{\bibfnamefont{M.~B.} \bibnamefont{Knickelbein}},
  \bibinfo{journal}{J. Chem. Phys.} \textbf{\bibinfo{volume}{125}},
  \bibinfo{pages}{044308} (\bibinfo{year}{2006}).

\bibitem[{\citenamefont{Douglass et~al.}(1993)\citenamefont{Douglass, Cox,
  Bucher, and Bloomfield}}]{douglass1993}
\bibinfo{author}{\bibfnamefont{D.~C.} \bibnamefont{Douglass}},
  \bibinfo{author}{\bibfnamefont{A.~J.} \bibnamefont{Cox}},
  \bibinfo{author}{\bibfnamefont{J.~P.} \bibnamefont{Bucher}},
  \bibnamefont{and} \bibinfo{author}{\bibfnamefont{L.~A.}
  \bibnamefont{Bloomfield}}, \bibinfo{journal}{Phys. Rev. B}
  \textbf{\bibinfo{volume}{47}}, \bibinfo{pages}{12874} (\bibinfo{year}{1993}).

\bibitem[{\citenamefont{de~Heer et~al.}(1990)\citenamefont{de~Heer, Milani, and
  Chatelain}}]{deheer1990}
\bibinfo{author}{\bibfnamefont{W.~A.} \bibnamefont{de~Heer}},
  \bibinfo{author}{\bibfnamefont{P.}~\bibnamefont{Milani}}, \bibnamefont{and}
  \bibinfo{author}{\bibfnamefont{A.}~\bibnamefont{Chatelain}},
  \bibinfo{journal}{Phys. Rev. Lett.} \textbf{\bibinfo{volume}{65}},
  \bibinfo{pages}{488} (\bibinfo{year}{1990}).

\bibitem[{\citenamefont{Cox et~al.}(1985)\citenamefont{Cox, Trevor, Whetten,
  Rohlfing, and Kaldor}}]{cox1985}
\bibinfo{author}{\bibfnamefont{D.~M.} \bibnamefont{Cox}},
  \bibinfo{author}{\bibfnamefont{D.~J.} \bibnamefont{Trevor}},
  \bibinfo{author}{\bibfnamefont{R.~L.} \bibnamefont{Whetten}},
  \bibinfo{author}{\bibfnamefont{E.~A.} \bibnamefont{Rohlfing}},
  \bibnamefont{and} \bibinfo{author}{\bibfnamefont{A.}~\bibnamefont{Kaldor}},
  \bibinfo{journal}{Phys. Rev. B} \textbf{\bibinfo{volume}{32}},
  \bibinfo{pages}{7290} (\bibinfo{year}{1985}).

\bibitem[{\citenamefont{Amirav and Navon}(1981)}]{Amirav1981}
\bibinfo{author}{\bibfnamefont{A.}~\bibnamefont{Amirav}} \bibnamefont{and}
  \bibinfo{author}{\bibfnamefont{G.}~\bibnamefont{Navon}},
  \bibinfo{journal}{Phys. Rev. Lett.} \textbf{\bibinfo{volume}{47}},
  \bibinfo{pages}{906} (\bibinfo{year}{1981}).

\bibitem[{\citenamefont{Kuebler et~al.}(1988)\citenamefont{Kuebler, Robin,
  Yang, Gedanken, and Herrick}}]{Kuebler1988}
\bibinfo{author}{\bibfnamefont{N.~A.} \bibnamefont{Kuebler}},
  \bibinfo{author}{\bibfnamefont{M.~B.} \bibnamefont{Robin}},
  \bibinfo{author}{\bibfnamefont{J.~J.} \bibnamefont{Yang}},
  \bibinfo{author}{\bibfnamefont{A.}~\bibnamefont{Gedanken}}, \bibnamefont{and}
  \bibinfo{author}{\bibfnamefont{D.~R.} \bibnamefont{Herrick}},
  \bibinfo{journal}{Phys. Rev. A} \textbf{\bibinfo{volume}{38}},
  \bibinfo{pages}{737} (\bibinfo{year}{1988}).

\bibitem[{\citenamefont{Khanna and Linderoth}(1991)}]{khanna1991}
\bibinfo{author}{\bibfnamefont{S.~N.} \bibnamefont{Khanna}} \bibnamefont{and}
  \bibinfo{author}{\bibfnamefont{S.}~\bibnamefont{Linderoth}},
  \bibinfo{journal}{Phys. Rev. Lett.} \textbf{\bibinfo{volume}{67}},
  \bibinfo{pages}{742} (\bibinfo{year}{1991}).

\bibitem[{\citenamefont{Knickelbein}(2004)}]{knickelbein2004}
\bibinfo{author}{\bibfnamefont{M.~B.} \bibnamefont{Knickelbein}},
  \bibinfo{journal}{J. Chem. Phys.} \textbf{\bibinfo{volume}{121}},
  \bibinfo{pages}{5281} (\bibinfo{year}{2004}).

\bibitem[{\citenamefont{McColm}(1964)}]{mccolm1964}
\bibinfo{author}{\bibfnamefont{D.}~\bibnamefont{McColm}},
  \bibinfo{journal}{Rev. Sci. Instrum.} \textbf{\bibinfo{volume}{37}},
  \bibinfo{pages}{1115} (\bibinfo{year}{1964}).

\bibitem[{\citenamefont{Li and Gu}(1993)}]{li1993}
\bibinfo{author}{\bibfnamefont{Z.~Q.} \bibnamefont{Li}} \bibnamefont{and}
  \bibinfo{author}{\bibfnamefont{B.~L.} \bibnamefont{Gu}},
  \bibinfo{journal}{Phys. Rev. B} \textbf{\bibinfo{volume}{47}},
  \bibinfo{pages}{13611} (\bibinfo{year}{1993}).

\bibitem[{\citenamefont{Guirado-Lopez et~al.}(2003)\citenamefont{Guirado-Lopez,
  Dorantes-Davila, and Pastor}}]{guirado2003}
\bibinfo{author}{\bibfnamefont{R.~A.} \bibnamefont{Guirado-Lopez}},
  \bibinfo{author}{\bibfnamefont{J.}~\bibnamefont{Dorantes-Davila}},
  \bibnamefont{and} \bibinfo{author}{\bibfnamefont{G.~M.}
  \bibnamefont{Pastor}}, \bibinfo{journal}{Phys. Rev. Lett}
  \textbf{\bibinfo{volume}{90}}, \bibinfo{pages}{226402}
  (\bibinfo{year}{2003}).

\bibitem[{\citenamefont{Gambardella et~al.}(2003)\citenamefont{Gambardella,
  Rusponi, Veronese, Dhesi, Grazioli, Dallmeyer, Cabria, Zeller, Dederichs,
  Kern et~al.}}]{gambardella2003}
\bibinfo{author}{\bibfnamefont{P.}~\bibnamefont{Gambardella}},
  \bibinfo{author}{\bibfnamefont{S.}~\bibnamefont{Rusponi}},
  \bibinfo{author}{\bibfnamefont{M.}~\bibnamefont{Veronese}},
  \bibinfo{author}{\bibfnamefont{S.~S.} \bibnamefont{Dhesi}},
  \bibinfo{author}{\bibfnamefont{C.}~\bibnamefont{Grazioli}},
  \bibinfo{author}{\bibfnamefont{A.}~\bibnamefont{Dallmeyer}},
  \bibinfo{author}{\bibfnamefont{I.}~\bibnamefont{Cabria}},
  \bibinfo{author}{\bibfnamefont{R.}~\bibnamefont{Zeller}},
  \bibinfo{author}{\bibfnamefont{P.~H.} \bibnamefont{Dederichs}},
  \bibinfo{author}{\bibfnamefont{K.}~\bibnamefont{Kern}}, \bibnamefont{et~al.},
  \bibinfo{journal}{Science} \textbf{\bibinfo{volume}{300}},
  \bibinfo{pages}{1130} (\bibinfo{year}{2003}).

\bibitem[{\citenamefont{de~Heer}()}]{electra2006}
\bibinfo{author}{\bibfnamefont{W.~A.} \bibnamefont{de~Heer}},
  \bibinfo{note}{see: http://www.physics.gatech.edu/npeg/co.html}.

\bibitem[{\citenamefont{Rodriguez-Lopez
  et~al.}(2003)\citenamefont{Rodriguez-Lopez, Aguilera-Granja, Michaelian, and
  Vega}}]{rodriguez2003}
\bibinfo{author}{\bibfnamefont{J.~L.} \bibnamefont{Rodriguez-Lopez}},
  \bibinfo{author}{\bibfnamefont{F.}~\bibnamefont{Aguilera-Granja}},
  \bibinfo{author}{\bibfnamefont{K.}~\bibnamefont{Michaelian}},
  \bibnamefont{and} \bibinfo{author}{\bibfnamefont{A.}~\bibnamefont{Vega}},
  \bibinfo{journal}{Phys. Rev. B} \textbf{\bibinfo{volume}{67}},
  \bibinfo{pages}{174413} (\bibinfo{year}{2003}).

\bibitem[{\citenamefont{Guevara et~al.}(1997)\citenamefont{Guevara, Parisi,
  Llois, and Weissmann}}]{guevara1997}
\bibinfo{author}{\bibfnamefont{J.}~\bibnamefont{Guevara}},
  \bibinfo{author}{\bibfnamefont{F.}~\bibnamefont{Parisi}},
  \bibinfo{author}{\bibfnamefont{A.~M.} \bibnamefont{Llois}}, \bibnamefont{and}
  \bibinfo{author}{\bibfnamefont{M.}~\bibnamefont{Weissmann}},
  \bibinfo{journal}{Phys. Rev. B} \textbf{\bibinfo{volume}{55}},
  \bibinfo{pages}{13283} (\bibinfo{year}{1997}).

\bibitem[{\citenamefont{Miura et~al.}(1994)\citenamefont{Miura, Kimura, and
  Imanaga}}]{miura1994}
\bibinfo{author}{\bibfnamefont{K.}~\bibnamefont{Miura}},
  \bibinfo{author}{\bibfnamefont{H.}~\bibnamefont{Kimura}}, \bibnamefont{and}
  \bibinfo{author}{\bibfnamefont{S.}~\bibnamefont{Imanaga}},
  \bibinfo{journal}{Phys. Rev. B} \textbf{\bibinfo{volume}{50}},
  \bibinfo{pages}{10335} (\bibinfo{year}{1994}).

\bibitem[{\citenamefont{Andriotis and Menon}(1998)}]{andriotis1998}
\bibinfo{author}{\bibfnamefont{A.~N.} \bibnamefont{Andriotis}}
  \bibnamefont{and} \bibinfo{author}{\bibfnamefont{M.}~\bibnamefont{Menon}},
  \bibinfo{journal}{Phys. Rev. B} \textbf{\bibinfo{volume}{57}},
  \bibinfo{pages}{10069} (\bibinfo{year}{1998}).

\bibitem[{\citenamefont{Yang and Knickelbein}(1990)}]{yang1990}
\bibinfo{author}{\bibfnamefont{S.}~\bibnamefont{Yang}} \bibnamefont{and}
  \bibinfo{author}{\bibfnamefont{M.}~\bibnamefont{Knickelbein}},
  \bibinfo{journal}{J. Chem. Phys.} \textbf{\bibinfo{volume}{93}},
  \bibinfo{pages}{1533} (\bibinfo{year}{1990}).

\bibitem[{\citenamefont{Parks et~al.}(1993)\citenamefont{Parks, Klots, Winter,
  and Riley}}]{parks1993}
\bibinfo{author}{\bibfnamefont{E.~K.} \bibnamefont{Parks}},
  \bibinfo{author}{\bibfnamefont{T.~D.} \bibnamefont{Klots}},
  \bibinfo{author}{\bibfnamefont{B.~J.} \bibnamefont{Winter}},
  \bibnamefont{and} \bibinfo{author}{\bibfnamefont{S.~J.} \bibnamefont{Riley}},
  \bibinfo{journal}{J. Chem. Phys.} \textbf{\bibinfo{volume}{99}},
  \bibinfo{pages}{5831} (\bibinfo{year}{1993}).

\bibitem[{\citenamefont{Payne et~al.}(2006)\citenamefont{Payne, Jiang, and
  Bloomfield}}]{payne2006}
\bibinfo{author}{\bibfnamefont{F.~W.} \bibnamefont{Payne}},
  \bibinfo{author}{\bibfnamefont{W.}~\bibnamefont{Jiang}}, \bibnamefont{and}
  \bibinfo{author}{\bibfnamefont{L.~A.} \bibnamefont{Bloomfield}},
  \bibinfo{journal}{Phys. Rev. Lett.} \textbf{\bibinfo{volume}{97}},
  \bibinfo{pages}{193401} (\bibinfo{year}{2006}).

\bibitem[{\citenamefont{Xie and Blackman}(2003)}]{xie2003}
\bibinfo{author}{\bibfnamefont{Y.}~\bibnamefont{Xie}} \bibnamefont{and}
  \bibinfo{author}{\bibfnamefont{J.}~\bibnamefont{Blackman}},
  \bibinfo{journal}{J. Phys.: Condens. Matter} \textbf{\bibinfo{volume}{15}},
  \bibinfo{pages}{L615} (\bibinfo{year}{2003}).

\bibitem[{\citenamefont{Fujima and Yamaguchi}(1996)}]{fujima1996}
\bibinfo{author}{\bibfnamefont{N.}~\bibnamefont{Fujima}} \bibnamefont{and}
  \bibinfo{author}{\bibfnamefont{T.}~\bibnamefont{Yamaguchi}},
  \bibinfo{journal}{Phys. Rev. B} \textbf{\bibinfo{volume}{54}},
  \bibinfo{pages}{26} (\bibinfo{year}{1996}).

\bibitem[{\citenamefont{Pellarin et~al.}(1994)\citenamefont{Pellarin,
  Baguenard, Vialle, Lerme, Broyer, Miller, and Perez}}]{pellarin1994}
\bibinfo{author}{\bibfnamefont{M.}~\bibnamefont{Pellarin}},
  \bibinfo{author}{\bibfnamefont{B.}~\bibnamefont{Baguenard}},
  \bibinfo{author}{\bibfnamefont{J.~L.} \bibnamefont{Vialle}},
  \bibinfo{author}{\bibfnamefont{J.}~\bibnamefont{Lerme}},
  \bibinfo{author}{\bibfnamefont{M.}~\bibnamefont{Broyer}},
  \bibinfo{author}{\bibfnamefont{J.}~\bibnamefont{Miller}}, \bibnamefont{and}
  \bibinfo{author}{\bibfnamefont{A.}~\bibnamefont{Perez}},
  \bibinfo{journal}{Chem. Phys. Lett.} \textbf{\bibinfo{volume}{217}},
  \bibinfo{pages}{349} (\bibinfo{year}{1994}).

\bibitem[{\citenamefont{Tiago et~al.}(2006)\citenamefont{Tiago, Zhou, Alemany,
  Saad, and Chelikowsky}}]{tiago2006}
\bibinfo{author}{\bibfnamefont{M.~L.} \bibnamefont{Tiago}},
  \bibinfo{author}{\bibfnamefont{Y.}~\bibnamefont{Zhou}},
  \bibinfo{author}{\bibfnamefont{M.~M.~G.} \bibnamefont{Alemany}},
  \bibinfo{author}{\bibfnamefont{Y.}~\bibnamefont{Saad}}, \bibnamefont{and}
  \bibinfo{author}{\bibfnamefont{J.~R.} \bibnamefont{Chelikowsky}},
  \bibinfo{journal}{Phys. Rev. Lett.} \textbf{\bibinfo{volume}{97}},
  \bibinfo{pages}{147201} (\bibinfo{year}{2006}).

\bibitem[{\citenamefont{Hamamoto et~al.}(2000)\citenamefont{Hamamoto, Onishi,
  and Bertsch}}]{hamamoto2000}
\bibinfo{author}{\bibfnamefont{N.}~\bibnamefont{Hamamoto}},
  \bibinfo{author}{\bibfnamefont{N.}~\bibnamefont{Onishi}}, \bibnamefont{and}
  \bibinfo{author}{\bibfnamefont{G.}~\bibnamefont{Bertsch}},
  \bibinfo{journal}{Phys. Rev. B} \textbf{\bibinfo{volume}{61}},
  \bibinfo{pages}{1336} (\bibinfo{year}{2000}).

\bibitem[{\citenamefont{Visuthikraisee and Bertsch}(1996)}]{visuthikraisee1996}
\bibinfo{author}{\bibfnamefont{V.}~\bibnamefont{Visuthikraisee}}
  \bibnamefont{and} \bibinfo{author}{\bibfnamefont{G.~F.}
  \bibnamefont{Bertsch}}, \bibinfo{journal}{Phys. Rev. A}
  \textbf{\bibinfo{volume}{54}}, \bibinfo{pages}{5104} (\bibinfo{year}{1996}).

\bibitem[{\citenamefont{Retort}(2003)}]{felix2003}
\bibinfo{author}{\bibfnamefont{F.}~\bibnamefont{Retort}},
  \bibinfo{journal}{Birkhause Verlag, Basel} p. \bibinfo{pages}{195}
  (\bibinfo{year}{2003}).

\bibitem[{\citenamefont{Bustamante et~al.}(2005)\citenamefont{Bustamante,
  Liphardt, and Ritort}}]{bustamante2005}
\bibinfo{author}{\bibfnamefont{C.}~\bibnamefont{Bustamante}},
  \bibinfo{author}{\bibfnamefont{J.}~\bibnamefont{Liphardt}}, \bibnamefont{and}
  \bibinfo{author}{\bibfnamefont{F.}~\bibnamefont{Ritort}},
  \bibinfo{journal}{Physics Today} \textbf{\bibinfo{volume}{58}},
  \bibinfo{pages}{43} (\bibinfo{year}{2005}).

\bibitem[{\citenamefont{Garnier and Ciliberto}(2005)}]{garnier2005}
\bibinfo{author}{\bibfnamefont{N.}~\bibnamefont{Garnier}} \bibnamefont{and}
  \bibinfo{author}{\bibfnamefont{S.}~\bibnamefont{Ciliberto}},
  \bibinfo{journal}{Phys. Rev. E} \textbf{\bibinfo{volume}{71}},
  \bibinfo{pages}{060101(R)} (\bibinfo{year}{2005}).

\bibitem[{\citenamefont{Amirav et~al.}(1980)\citenamefont{Amirav, Even, and
  Jortner}}]{amirav1980}
\bibinfo{author}{\bibfnamefont{A.}~\bibnamefont{Amirav}},
  \bibinfo{author}{\bibfnamefont{U.}~\bibnamefont{Even}}, \bibnamefont{and}
  \bibinfo{author}{\bibfnamefont{J.}~\bibnamefont{Jortner}},
  \bibinfo{journal}{Chem. Phys.} \textbf{\bibinfo{volume}{51}},
  \bibinfo{pages}{31} (\bibinfo{year}{1980}).

\bibitem[{\citenamefont{Geusic et~al.}(1985)\citenamefont{Geusic, Morse,
  O'Brien, and Smalley}}]{geusic1985}
\bibinfo{author}{\bibfnamefont{M.~E.} \bibnamefont{Geusic}},
  \bibinfo{author}{\bibfnamefont{M.~D.} \bibnamefont{Morse}},
  \bibinfo{author}{\bibfnamefont{S.~C.} \bibnamefont{O'Brien}},
  \bibnamefont{and} \bibinfo{author}{\bibfnamefont{R.~E.}
  \bibnamefont{Smalley}}, \bibinfo{journal}{Rev. Sci. Instrum.}
  \textbf{\bibinfo{volume}{56}}, \bibinfo{pages}{2123} (\bibinfo{year}{1985}).

\bibitem[{\citenamefont{Wallraff et~al.}(1987)\citenamefont{Wallraff, Yamada,
  and Winnewisser}}]{wallraff1987}
\bibinfo{author}{\bibfnamefont{P.}~\bibnamefont{Wallraff}},
  \bibinfo{author}{\bibfnamefont{K.~M.~T.} \bibnamefont{Yamada}},
  \bibnamefont{and}
  \bibinfo{author}{\bibfnamefont{G.}~\bibnamefont{Winnewisser}},
  \bibinfo{journal}{J. Mole. Spectr.} \textbf{\bibinfo{volume}{126}},
  \bibinfo{pages}{78} (\bibinfo{year}{1987}).

\bibitem[{\citenamefont{Collings et~al.}(1993)\citenamefont{Collings, Amrein,
  Rayner, and Hackett}}]{collings1993}
\bibinfo{author}{\bibfnamefont{B.~A.} \bibnamefont{Collings}},
  \bibinfo{author}{\bibfnamefont{A.~H.} \bibnamefont{Amrein}},
  \bibinfo{author}{\bibfnamefont{D.~M.} \bibnamefont{Rayner}},
  \bibnamefont{and} \bibinfo{author}{\bibfnamefont{P.~A.}
  \bibnamefont{Hackett}}, \bibinfo{journal}{J. Chem. Phys.}
  \textbf{\bibinfo{volume}{99}}, \bibinfo{pages}{4174} (\bibinfo{year}{1993}).

\bibitem[{\citenamefont{Frenken and van~der Veen}(1985)}]{frenken1985}
\bibinfo{author}{\bibfnamefont{J.~W.~M.} \bibnamefont{Frenken}}
  \bibnamefont{and} \bibinfo{author}{\bibfnamefont{J.~F.} \bibnamefont{van~der
  Veen}}, \bibinfo{journal}{Phys. Rev. Lett.} \textbf{\bibinfo{volume}{54}},
  \bibinfo{pages}{134} (\bibinfo{year}{1985}).

\bibitem[{\citenamefont{Bohr}(2001)}]{bohr2001}
\bibinfo{author}{\bibfnamefont{J.}~\bibnamefont{Bohr}}, \bibinfo{journal}{Int.
  J. Quant. Chem.} \textbf{\bibinfo{volume}{84}}, \bibinfo{pages}{249}
  (\bibinfo{year}{2001}).

\bibitem[{\citenamefont{Bachels et~al.}(2000)\citenamefont{Bachels, Guntherodt,
  and Schafer}}]{bachels2000}
\bibinfo{author}{\bibfnamefont{T.}~\bibnamefont{Bachels}},
  \bibinfo{author}{\bibfnamefont{H.-J.} \bibnamefont{Guntherodt}},
  \bibnamefont{and} \bibinfo{author}{\bibfnamefont{R.}~\bibnamefont{Schafer}},
  \bibinfo{journal}{Phys. Rev. Lett.} \textbf{\bibinfo{volume}{85}},
  \bibinfo{pages}{1250} (\bibinfo{year}{2000}).

\bibitem[{\citenamefont{Breaux et~al.}(2005)\citenamefont{Breaux, Neal, Cao,
  and Jarrold}}]{breaux2005}
\bibinfo{author}{\bibfnamefont{G.~A.} \bibnamefont{Breaux}},
  \bibinfo{author}{\bibfnamefont{C.~M.} \bibnamefont{Neal}},
  \bibinfo{author}{\bibfnamefont{B.}~\bibnamefont{Cao}}, \bibnamefont{and}
  \bibinfo{author}{\bibfnamefont{M.~F.} \bibnamefont{Jarrold}},
  \bibinfo{journal}{Phys. Rev. Lett.} \textbf{\bibinfo{volume}{94}},
  \bibinfo{pages}{173401} (\bibinfo{year}{2005}).

\bibitem[{\citenamefont{Zhao et~al.}(1996)\citenamefont{Zhao, de~Beer, Xu,
  Taylor, and Neumark}}]{zhao1996}
\bibinfo{author}{\bibfnamefont{Y.}~\bibnamefont{Zhao}},
  \bibinfo{author}{\bibfnamefont{E.}~\bibnamefont{de~Beer}},
  \bibinfo{author}{\bibfnamefont{C.}~\bibnamefont{Xu}},
  \bibinfo{author}{\bibfnamefont{T.~R.} \bibnamefont{Taylor}},
  \bibnamefont{and} \bibinfo{author}{\bibfnamefont{D.~M.}
  \bibnamefont{Neumark}}, \bibinfo{journal}{J. Chem. Phys.}
  \textbf{\bibinfo{volume}{105}}, \bibinfo{pages}{4905} (\bibinfo{year}{1996}).

\bibitem[{\citenamefont{Xu et~al.}(1997)\citenamefont{Xu, Burton, Taylor, and
  Neumark}}]{xu1997}
\bibinfo{author}{\bibfnamefont{C.}~\bibnamefont{Xu}},
  \bibinfo{author}{\bibfnamefont{G.~R.} \bibnamefont{Burton}},
  \bibinfo{author}{\bibfnamefont{T.~R.} \bibnamefont{Taylor}},
  \bibnamefont{and} \bibinfo{author}{\bibfnamefont{D.~M.}
  \bibnamefont{Neumark}}, \bibinfo{journal}{J. Chem. Phys.}
  \textbf{\bibinfo{volume}{107}}, \bibinfo{pages}{3428} (\bibinfo{year}{1997}).

\bibitem[{\citenamefont{Guirado-Lopez}(2001)}]{guirado2001}
\bibinfo{author}{\bibfnamefont{R.}~\bibnamefont{Guirado-Lopez}},
  \bibinfo{journal}{Phys. Rev. B} \textbf{\bibinfo{volume}{63}},
  \bibinfo{pages}{174420} (\bibinfo{year}{2001}).

\bibitem[{\citenamefont{Xie and Blackman}(2004)}]{xie2004}
\bibinfo{author}{\bibfnamefont{Y.}~\bibnamefont{Xie}} \bibnamefont{and}
  \bibinfo{author}{\bibfnamefont{J.~A.} \bibnamefont{Blackman}},
  \bibinfo{journal}{J. Phys.: Condens. Matter} \textbf{\bibinfo{volume}{16}},
  \bibinfo{pages}{3163} (\bibinfo{year}{2004}).

\end{thebibliography}

\end{document}